\documentclass[a4paper,fleqn]{cas-dc}

\usepackage[authoryear]{natbib}
\usepackage{amsmath,amssymb,amsfonts}
\usepackage{algorithmic}
\usepackage{graphicx}
\usepackage{textcomp}
\usepackage{xcolor}
\usepackage{stfloats}
\usepackage{ragged2e}
\usepackage{balance}
\usepackage{colortbl}
\usepackage{multirow}
\usepackage{balance}
\usepackage[figuresright]{rotating}
\usepackage{rotfloat}
\usepackage{color}

\usepackage{CJKutf8}
\usepackage{framed}

\usepackage[utf8]{inputenc}
\usepackage{tcolorbox}
\usepackage{enumitem}

\usepackage{hyperref}
\usepackage{hyperref}

\hypersetup{
    colorlinks=true,
    linkcolor=blue,
    anchorcolor=blue,
    citecolor=blue
}
\usepackage{booktabs}
\usepackage{pifont}
\newcommand{\xmark}{\ding{55}} 

\usepackage{pgfplots}
\usepackage{pgfplotstable}
\usepackage{graphicx}
\pgfplotsset{compat=1.18}

\usepackage{enumitem}  

\def\tsc#1{\csdef{#1}{\textsc{\lowercase{#1}}\xspace}}
\tsc{WGM}
\tsc{QE}
\tsc{EP}
\tsc{PMS}
\tsc{BEC}
\tsc{DE}

\begin{document}
\begin{sloppypar}

\let\WriteBookmarks\relax
\def\floatpagepagefraction{1}
\def\textpagefraction{.001}

\shorttitle{CoRaCMG: Contextual Retrieval-Augmented Framework for Commit Message Generation}
\shortauthors{X Bo et al.}
\title [mode = title]{CoRaCMG: Contextual Retrieval-Augmented Framework for Commit Message Generation}

\author[1]{Bo Xiong}
\fnmark[1]
\ead[1]{yueshaomoon_@whu.edu.cn}
\credit{Conceptualization, Methodology, Investigation, Data curation, Software, Writing - Original draft preparation}

\author[1]{Linghao Zhang}
\fnmark[1] 
\ead[2]{starryzhang@whu.edu.cn}
\credit{Conceptualization, Methodology, Investigation, Data curation, Software, Writing - Original draft preparation}

\author[1]{Zongen Ren}
\ead[2]{zongnren@whu.edu.cn}
\credit{Investigation, Data curation, Software, Writing - Original draft preparation}

\author[1]{Chong Wang}
\cormark[1]
\ead[3]{cwang@whu.edu.cn}
\credit{Conceptualization, Methodology, Investigation, Writing - Original draft preparation}

\author[1]{Peng Liang}
\cormark[1] 
\ead[4]{liangp@whu.edu.cn}
\credit{Conceptualization, Methodology, Investigation, Writing - Original draft preparation}

\affiliation[1]{organization={School of Computer Science, Wuhan University},
             city={Wuhan},
             country={China}}
\fntext[1]{These authors contributed equally to this work.}
\cortext[1]{Corresponding authors.}








\begin{abstract}
\textbf{}\textbf{Context}: In software development and maintenance, commit messages play a key role in documenting the intent behind code changes, supporting collaboration, review, and maintenance. However, they are often low-quality, vague, or incomplete, limiting their usefulness. Commit Message Generation (CMG) aims to automatically generate descriptive commit messages from code diffs to reduce developers' effort and improve message quality. Although recent advances in Large Language Models (LLMs) have shown promise in automating CMG, their performance remains limited.\\
\textbf{Objective}: This paper aims to enhance CMG performance by retrieving similar diff-message pairs to guide LLMs to generate commit messages that are more precise and informative.\\
\textbf{Method}: We proposed CoRaCMG, a \textbf{Co}ntextual \textbf{R}etrieval-\textbf{a}ugmented framework for \textbf{C}ommit \textbf{M}essage \textbf{G}eneration equipped with a hybrid retrieval mechanism, structured in three phases: (1) Retrieve: retrieving the similar diff-message pairs; (2) Augment: combining them with the query diff into a structured prompt; and (3) Generate: generating commit messages corresponding to the query diff via LLMs. CoRaCMG enables LLMs to learn project-specific terminologies and writing styles from the retrieved diff-message pairs.\\
\textbf{Results}: We evaluated CoRaCMG across multiple LLMs (e.g., GPT, DeepSeek, and Qwen) and compared its performance against SOTA baselines. Experimental results show that CoRaCMG achieves performance comparable to or superior to these SOTA baselines and significantly boosts LLM performance across four metrics (BLEU, Rouge-L, METEOR, and CIDEr). Specifically, DeepSeek-R1 achieves relative improvements of 76\% in BLEU and 71\% in CIDEr when augmented with a single retrieved example pair. After incorporating the single example pair, GPT-4o achieves the highest improvement rate, with BLEU increasing by 89\%. Moreover, performance gains plateau after more than three examples are used, indicating diminishing returns. Further analysis indicates that the improvements stem from the model's capability to capture the terminologies and writing styles from the retrieved example pairs. Moreover, our experiments demonstrate that CoRaCMG generalizes effectively across multiple programming languages, including Java, Python, and C++, while human evaluation also confirms its ability to produce commit messages with higher quality than baseline approaches.\\
\textbf{Conclusions}: CoRaCMG provides an effective framework for the CMG task, supported by comprehensive evaluations that demonstrate significant improvements in the quality of generated commit messages.
\end{abstract}


\maketitle

\section{Introduction}
\label{chap:intro}
In software development and maintenance, the Git version control system has been widely used to store and share code. Within Git, commit messages describe and document the code changes made in a commit. These messages record alterations to the source code, help developers understand the changes in the code, and promote efficient collaboration. Therefore, the commit message serves as one of the critical pieces of textual information in the software engineering life-cycle. However, writing commit messages is time-consuming and laborious for developers. Many developers find them tedious and are not motivated to write~\citep{maalej2010can}. As a result, in real-world development settings, the overall quality of commit messages is often suboptimal. As a recent work reported by \cite{tian2022makes}, on average 44\% of commit messages did not reach the desired quality, indicating a lack of essential information and the struggle to convey critical details about what the commit did and why.

CMG task aims to take the differences between two versions of code as input, typically in the form of a code diff file \texttt{.diff} generated by Git, aka code change, and then generates the corresponding commit message. Initially, rule-based approaches were used in CMG~\citep{buse2010automatically,vasquez2015changescribe,shen2016automatic}, where predefined rules or templates were utilized for generation. Some retrieval-based approaches leverage information retrieval (IR) techniques to suggest commit messages from similar code diff~\citep{liu2018neural,hoang2020cc2vec}. With the rapid development of deep learning, many learning-based methods have recently emerged~\citep{dong2022fira,jiang2017automatically,liu2019generating,loyola2017neural,xu2019commit}. These methods treat CMG as a neural machine translation task, in which code diff is the input of a neural network model to generate commit messages as output. In addition, there are some hybrid approaches~\citep{liu2022atom,shi2022race,wang2021CoRec} that leverage both IR and deep learning techniques. With the recent advances in LLMs, these models have been increasingly applied to both natural language processing and code-related tasks. Several studies~\citep {zhang2024automatic,zhang2024using} have begun exploring their potential on the CMG task and have reported promising prospects.

Additionally, several publicly available datasets~\citep{liu2018neural,jiang2017automatically,xu2019commit,wang2021CoRec} have been widely adopted in recent research. For instance, \cite{jiang2017automatically} constructed a dataset exclusively comprising commit messages from Java projects, which has been reused in numerous subsequent studies~\citep{buse2010automatically,dong2022fira,zhang2024using,xu2019commit}. Table \ref{tab:datasets} presents an overview of several commonly used commit message datasets along with their associated statistical information. These datasets are typically constructed by crawling commit records from open-source projects hosted on GitHub. However, they often suffer from some issues related to data quality and the standardization of their construction processes.

A study by \cite{zhang2024using} revealed a notable phenomenon: in a human evaluation experiment, participants were tasked with selecting the optimal commit message from a set of generated messages (including the reference message authored by developers) for a given commit. The experiment utilized a widely adopted commit message dataset \citep{xu2019commit}. The results indicated that across 366 commit samples, only 13.1\% of the samples had the developer-authored reference commit message selected as the optimal choice. This finding underscores the inherent quality issues within these datasets and highlights their potential impact on standardized evaluation protocols for the CMG task. 
Given the limitations in quality and standardization of existing datasets, we argue for the necessity of constructing a new, high-quality dataset.

\begin{figure*}[pos=h]
    \centering
    \includegraphics[width=0.8\linewidth]{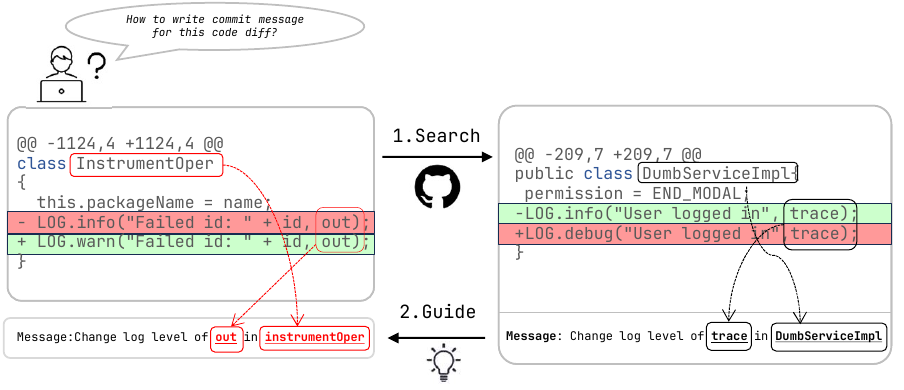}
    \caption{A motivating scenario of how developers write a commit message by referring to a similar diff-message pair}
    \label{fig:motivation}
\end{figure*}

However, directly applying LLMs to CMG still faces several challenges. First, the generated messages may overlook project-specific writing styles. Second, the input context is often limited and lacks a global understanding of the codebase, which can result in missing critical information necessary for accurate generation. Most importantly, existing models have not yet effectively leveraged a key characteristic in software engineering: historical commit messages often contain the terminology of human-written commit messages and the writing styles.
Against this backdrop, Retrieval Augmented Generation (RAG) offers a promising direction for CMG ~\citep{Gao2023RAGSurvey, ram2023incontext, lewis2020retrieval, wang2023rapRetrieval}. By retrieving external knowledge to enrich the input with relevant information, RAG enhances the performance of generation models. Nevertheless, in the domain of CMG, existing work has yet to fully explore how to effectively integrate RAG with LLMs, nor has it thoroughly examined the potential performance gains from such integration.

Consider the typical scenario in Figure~\ref{fig:motivation}, where if a developer wants to write a good commit message, an efficient way is to search for and imitate a similar diff-message pair. That is, exemplary commit messages associated with similar code diffs can be of significant value, as they offer supplementary information that contributes to enhancing the performance of commit message generation. This manual ``search-and-imitate'' process illustrates that retrieving relevant historical context is crucial for producing high-quality commit messages. Inspired by this insight, we proposed a framework called CoRaCMG, a \textbf{Co}ntextual \textbf{R}etrieval-\textbf{a}ugmented framework for \textbf{C}ommit \textbf{M}essage \textbf{G}eneration equipped with a hybrid retrieval mechanism. Unlike the approaches that rely on computationally expensive and static fine-tuning—which struggles to adapt to new projects without retraining—CoRaCMG employs a training-free, dynamic retrieval mechanism to automate this ``search-and-imitate'' process. CoRaCMG allows LLMs to explicitly capture and leverage project-specific terminologies and writing styles directly from retrieved similar diff-message pairs. Furthermore, this mechanism of CoRaCMG exhibits cross-language generalizability, enabling high-quality commit message generation across diverse programming languages without the need for language-specific training. We conducted the experiments on a high-quality and comprehensive CMG dataset to ensure reliable evaluation of the proposed framework. Our experiment results show that this approach brings substantial performance gains compared to directly applying LLMs without retrieving similar diff–message pairs.

\textbf{Our approach}: 
CoRaCMG comprises three phases, i.e., Retrieve, Augment, and Generate, to generate commit messages for a given code diff. \textbf{Retrieve}: We design a hybrid retriever to retrieve the most relevant diff and its corresponding commit message from a comprehensive and high-quality source database. The retrieved diff-message pair will be used to guide the subsequent generation of the commit message. \textbf{Augment}: The query diff and retrieved diff-message pair need to be combined as the input of the models, that is, input augmentation. The augmenter concatenates them with predefined special tokens or fills them into a prompt template before passing the result to the generator for the next phase. \textbf{Generate}: The generator receives augmented inputs and generates a commit message corresponding to the query diff under the guidance of the retrieved example pairs. We conduct extensive experiments to evaluate the effectiveness of CoRaCMG across various LLMs.The results demonstrate that CoRaCMG enables models to effectively leverage additional contextual information, thereby improving the completeness of the generated commit messages.

The main \textbf{contributions} of our work can be summarized as follows, and the replication package is available at~\citep{replpack}:
\begin{itemize}   
    \item \textbf{An effective retrieval-augmented framework for CMG.} We propose CoRaCMG, a retrieval-augmented framework designed to enhance CMG. CoRaCMG effectively integrates advanced retrieval techniques with various generation models, thereby improving the quality and informativeness of the generated commit messages.
    
    \item \textbf{Comprehensive experimental evaluation of the proposed framework.} We conduct extensive experiments to assess the effectiveness of CoRaCMG. The evaluation results demonstrate that CoRaCMG consistently improves the performance of LLMs compared to using them directly for generation. Specifically, in terms of BLEU scores, CoRaCMG achieves an average improvement of 65\%.
\end{itemize}

The remainder of this paper is structured as follows: Section~\ref{RelatedWork} covers related work on CMG. Section~\ref{chap:ApacheCM} provides a detailed introduction to ApacheCM dataset. Section~\ref{chap:methodology} introduces our CoRaCMG framework and its three phases. Section~\ref{Experimental Setup} describes the experimental setup, including the dataset and model selection, while Section~\ref{chap: result} presents results and analysis. Finally, Section~\ref{chap:threats} discusses the threats to validity, followed by the conclusions in Section ~\ref{chap: conclusions}.

\section{Related Work}
\label{RelatedWork}
Various approaches for the CMG task have been proposed in recent years. According to the generating mechanisms of commit messages, these approaches can be categorized into retrieval-based, learning-based, and hybrid methods, as well as LLM-based methods.

\subsection{Retrieval-based Methods} 
NNGen ~\citep{liu2018neural} leverages information retrieval techniques to suggest commit messages from similar code diffs. To generate a commit message, NNGen calculates the cosine similarity between the target code diff and each code diff in the collected corpus. Then, the top-\textit{k} diff-message pairs are selected to compute the BLEU scores; the one with the highest score is regarded as the most similar code diff, and its commit message will be used as the target one. CC2Vec ~\citep{hoang2020cc2vec} learns a representation of code diff guided by their accompanying commit messages. Similar to the nearest neighbors approach, it computes the distance between code diff vectors and directly outputs the commit message of the closest CC2Vec vector. Retrieval-based methods have significant challenges as the corpus is limited and cannot cover all code diffs.

Retrieval-based methods have the advantages of being simple to implement, training-free, and offering interpretable outputs. However, it also presents certain limitations: (1) they heavily rely on the quality and diversity of the dataset; and (2) when the query diff lacks similar example pairs in the dataset, the quality of retrieved pairs drops significantly, thereby affecting the generation performance.

\subsection{Learning-based Methods}
Learning-based methods \citep{jiang2017automatically,liu2019generating,liu2018neural,loyola2017neural,xu2019commit,huang2025commit} typically employ deep learning techniques, treating CMG as a Neural Machine Translation (NMT) task. These methods learn how to generate commit messages by training deep neural network models on massive diff-message datasets collected from GitHub projects. CommitGen ~\citep{jiang2017automatically} is an early attempt to use NMT in CMG; it trained a recurrent neural network (RNN) encoder-decoder model using a corpus of diffs and human-written commit messages from the top 1k GitHub projects. CoDiSum ~\citep{xu2019commit} uses a multi-layer bidirectional gated recurrent unit (GRU) as its encoder part, which can better learn the representations of code changes. Moreover, the copying mechanism is used in the decoder part to mitigate the out-of-vocabulary (OOV) issue. PtrGNCMsg ~\citep{liu2019generating} is another NMT approach based on an improved sequence-to-sequence model with the pointer-generator network, which is an adapted version of the attention RNN encoder-decoder model. FIRA ~\citep{dong2022fira} first represents the code diffs with fine-grained graphs, which explicitly describe the code edit operations between the old version and the new version, and code tokens at different granularities. The hybrid architecture of transformer and GNN is adopted as the backbone of the model. FIRA outperforms other learning-based methods and can be considered the current SOTA approach of a single model. CCT5~\citep{Lin2023CCT5} is a code-diff-oriented pre-trained model built on the T5 architecture, which constructs a large-scale dataset CodeChangeNet and designs five code-change-oriented pre-training tasks to effectively capture the semantic information of code diffs. The recently proposed learning-based approach, CCGen~\citep{huang2025commit}, integrates a Transformer-based framework with a pre-trained CodeBERT encoder and introduces a novel method for identifying the ``core change'' in a commit - the most critical code modification. Core changes are determined using a CatBoost classifier based on features such as code modification, coupling, and semantics, enabling the model to focus on these core changes to generate more accurate commit messages, particularly for multi-class commits.

Learning-based methods can automatically learn the terminologies and writing styles of human-written commit messages from retrieved diff–message pairs. However, they are highly dependent on the scale and quality of the training dataset, have limited adaptability to other projects whose terminologies and writing styles are not covered in the dataset, and often incur high computational costs.
\subsection{Hybrid Methods}
Hybrid methods are a combination of retrieval-based methods and learning-based methods. ATOM ~\citep{liu2022atom} is a hybrid method containing three modules, a generation module encoding the structure of code diff using Abstract Syntax Tree (AST), a retrieval module retrieving the most similar message based on the text-similarity, and a hybrid ranking module selecting the best commit message from the ones generated by generation and retrieval modules. CoRec~\citep{wang2021CoRec} takes advantage of both IR and NMT, addressing the low-frequency word and exposure bias issue. It trained a context-aware encoder-decoder model. Given a diff for testing, the method retrieves the most similar diff from the training set and then uses it to guide the probability distribution for the final generated vocabulary. RACE~\citep{shi2022race} combines retrieval and generation techniques in a more integrated way but employs a trivial retriever and opts for training a specific model from scratch. COME~\citep{He2023COME} utilizes modification embedding based on edit distance to represent code changes in a fine-grained manner, combines a translation module with a retrieval module, and employs an SVM-based decision algorithm to evaluate the outputs of both modules simultaneously, dynamically selecting the higher-quality commit message as the final result. Whereas our proposed framework, CoRaCMG, incorporates a hybrid advanced retriever, and experimental results demonstrate its effectiveness. Furthermore, CoRaCMG is not limited to a specific model, but rather widely adopts various successful models as the generator, enabling more effective generation of commit messages and providing generality insights.

\subsection{LLM-based Methods}
In recent years, LLMs have demonstrated exceptional capabilities in text comprehension and generation. Several studies~\citep{zhang2024using,xiong2025C3GEN, huang2025commit} have investigated the application of LLMs to CMG, achieving impressive results.~\cite{zhang2024using} conducted a systematic experimental study on the application of LLMs to this task. Their findings indicate that, without additional training or fine-tuning, LLMs can generate more accurate commit messages through carefully designed prompts, surpassing traditional baseline methods across multiple objective metrics. Moreover, in human evaluations, developers exhibited greater acceptance of commit messages generated by LLMs, further underscoring their potential in this domain.

\cite{li2024only} proposed OMG, which reframes CMG as a knowledge-intensive reasoning task. OMG employs the ReAct (Reasoning + Acting)~\citep{yao2023react} prompting strategy to guide LLMs in dynamically invoking multiple external tools during generation, thereby acquiring more comprehensive software context. Evaluations demonstrate that OMG significantly outperforms state-of-the-art conventional methods, such as FIRA, across multiple human-assessed metrics. Furthermore, OMG surpasses developer-written commit messages in several dimensions, including rationality, comprehensiveness, and expressiveness.~\cite{kumar2025using} proposed a novel approach using LLMs to generate file-level and commit-level messages for large diffs. Employing prompt engineering and hyperparameter tuning, it enhances message quality for complex, multi-file changes, outperforming traditional methods in automated and human evaluations. ~\cite{xiong2025C3GEN} proposed C3Gen, a retrieval-enhanced framework that augments prompts with semantically relevant code snippets retrieved from the repository using Code Structure Graphs built via Tree-sitter, leading to more complete and informative commit messages in human evaluations.

LLM-based methods broaden the application of text generation in software engineering and demonstrate promising prospects in the CMG task.

\subsection{Conclusive Summary} 
While current studies on CMG have made promising progress, they have not thoroughly explored how to effectively integrate RAG with LLMs so as to provide LLMs with the repository-historical context, such as past code diffs and their corresponding messages. CoRaCMG addresses this limitation and demonstrates significant differences from prior approaches in both design philosophy and technical implementation. The key distinctions are summarized as follows:

First, in contrast to Retrieval-based methods (e.g., NNGen~\citep{liu2018neural}, CC2Vec~\citep{hoang2020cc2vec}), which generate commit messages by directly reusing an existing message from its similar diff, CoRaCMG adopts a generative approach. Retrieval-based methods suffer from inherent inaccuracy if the retrieved message is not a perfect match for the query diff, as they directly treat the retrieved message as the final output. Differently, CoRaCMG uses the retrieved message strictly as a ``reference exemplar'' for guidance rather than the final output. The LLM acts as a reasoning and generating engine that learns the writing style and terminology from the retrieved example and generates new message that accurately describes the current query diff.

Second, compared to Learning-based methods (e.g., CCT5~\citep{Lin2023CCT5}, FIRA~\citep{dong2022fira} ), which treat CMG as a translation task and rely heavily on computationally expensive training or fine-tuning on massive datasets, CoRaCMG is a training-free framework. Learning-based methods typically produce static models; once trained, they cannot adapt to a new project's specific writing style without retraining. CoRaCMG utilizes a dynamic retrieval mechanism that allows the model to adapt to any project immediately by simply accessing its historical diff-message pairs, thereby eliminating the high cost of training and time that static models face.

Third, distinct from Hybrid methods (e.g., CoRec~\citep{wang2021CoRec}, RACE~\citep{shi2022race} ) that combine retrieval with generation, CoRaCMG decouples the retrieval and generation phases into a flexible, modular framework. Previous hybrid methods often rely on complex, bespoke architectures (e.g., training specific encoders/decoders alongside retrieval modules) to ``fuse'' retrieval results. In contrast, CoRaCMG leverages the In-Context Learning capability of  LLMs. This makes CoRaCMG more generalizable and easier to deploy across different programming languages and projects without architecture modification.

Finally, regarding recent LLM-based methods, CoRaCMG addresses challenges different from those targeted by works such as ~\cite{kumar2025using}. While standard Zero-shot LLM approaches often generate generic messages that overlook project-specific conventions, ~\cite{kumar2025using} primarily focus on the challenge of ``Large Diffs'' (input length constraints) via a hierarchical approach. Our work, however, addresses the dimension of ``Writing Style and project-specific Terminology''. CoRaCMG employs RAG to ensure the generated message aligns with the historical norms of the repository (e.g., using correct abbreviations and sentence structures). While Kumar et al. solve the ``Length'' problem, CoRaCMG effectively solves the ``Writing Style and project-specific Terminology'' problem.

\section{ApacheCM}
\label{chap:ApacheCM}

\subsection{Existing Datasets and Their Limitations}
Several publicly available datasets have been widely adopted in recent research on CMG, including CommitGen \citep{jiang2017automatically}, NNGen \citep{liu2018neural}, CoDiSum \citep{xu2019commit}, MCMD \citep{wang2021CoRec}, etc. Taking CommitGen as an example, it contains commit messages only from Java projects and has been reused in \cite{buse2010automatically, dong2022fira, zhang2024using, xu2019commit}. The first four columns of  Table~\ref{tab:datasets} present a summary of the fundamental details pertaining to four commonly used commit message datasets mentioned above. The details of these four datasets are as follows.

\begin{table*}[pos=htbp]
\centering
\caption{ApacheCM v.s. Existing Datasets}
\label{tab:datasets}
\begin{tabular}{lcccccc}
\toprule
\textbf{Dataset} & \textbf{CommitGen} & \textbf{NNGen} & \textbf{CoDiSum} & \textbf{MCMD} & \textbf{ApacheCM} \\
\midrule
\#Train Set         & 26,208    & 22,112  & 75,000     & 1,800,000  & 249,830 \\
\#Validation Set    & 3,000     & 3,000   & 8,000      & 225,000    & 10,000 \\
\#Test Set          & 3,000     & 2,521   & 7,661      & 225,000    & 10,000 \\
\#Repositories      & 1,000     & 1,000   & 1,000      & 500        & 50 \\
Filtering Steps     & \xmark    & \xmark  & \checkmark & \checkmark & \checkmark \\
Deduplication       & \xmark    & \xmark  & \xmark     & \checkmark & \checkmark \\
Reproducible        & \xmark    & \xmark  & \xmark     & \xmark     & \checkmark \\
Complete Schema     & \xmark    & \xmark  & \xmark     & \xmark     & \checkmark \\
\rule{0pt}{1.9em}Languages Covered   & Java      & Java    & Java       & \makecell[l]{Java, C\#, C++,\\ Python, JavaScript} & \makecell[l]{Nine types of programming\\ languages listed in Table \ref{tab:lang_stats}} \\

\rule{0pt}{2.5em}Extracted Data Item & 
Diff, Message & 
Diff, Message & 
Diff, Message & 
\makecell[l]{Diff, Message,\\ RepoName, SHA, \\Timestamp} & 
\makecell[l]{Diff, Message, RepoName,\\ SHA, Author Name, Files,\\ Date, LoC} \\
\bottomrule
\end{tabular}
\end{table*}

\begin{itemize}
    \item \textbf{CommitGen} is the most used dataset for CMG. It is a Java-only and small-scale dataset comprising code-diff and commit messages from
    the top 1,000 Java repositories. It employs a Verb-Direct Object filter to improve message quality, but introduces issues such as trivial and bot-generated messages and duplicates.
    
    \item \textbf{NNGen} is a refined subset of CommitGen, by applying more sophisticated filtering rules. 
    Although the data quality of NNGen was improved by removing trivial or bot-generated messages, it still suffers from limitations similar to those of CommitGen.
    
    \item \textbf{CoDiSum} is another Java-only dataset based on the top 1,000 Java repositories. Among these four datasets, it is uniquely incorporates a file-level programming language filter. However, CoDiSum lacks other quality filtering strategies and results in a larger-scale dataset than those from similar data sources.
    
    \item \textbf{MCMD} is a large-scale, multi-language dataset comprising code commits from the top 500 GitHub repositories, spanning five programming languages listed in the ninth row of Table 1. However, it fails to implement a sequence length filter, thereby rendering the dataset excessively large and posing processing challenges for the majority of models.
\end{itemize}

The four datasets for CMG mentioned above have at least two limitations. First, most of these datasets are Java-only. They are plagued not only by data quality issues but also by a notable absence of standardization in their construction process. 
Among these, only CoDiSum and MCMD implement supplementary quality control filters, only MCMD undertakes deduplication efforts, and none of them currently offer reproducibility.
Second, these four datasets typically consisted of code diffs paired with commit messages and omitted critical contextual metadata such as repository names, commit SHA values, and timestamps. The lack of contextual information significantly limits their applicability in broader research scenarios. Taking CoDiSum as an example, it was collected from the top 1,000 Java repositories on GitHub, encompassing 90,661 pairs of <\texttt{diff}, \texttt{message}>. CoDiSum incorporates only minimal quality control measures and lacks standardized criteria for repository selection. Furthermore, it does not provide access to the raw data, making the collection process difficult to reproduce. More critically, each record in CoDiSum only includes the code diff and its corresponding commit message, omitting other essential metadata. These omissions significantly limit the use of these datasets in broader research.

Given the limitations of existing datasets, we base our experiments on the ApacheCM dataset~\citep{xiong2025C3GEN}, a high-quality and comprehensive dataset for CMG. As shown in the last column of Table~\ref{tab:datasets}, it covers high-quality projects in nine programming languages, and also provides comprehensive contextual metadata and ensures high data quality. Besides supporting our experiments in this paper, ApacheCM is potentially expected to serve as a standardized and reproducible benchmark for future research in CMG. The details of ApacheCM are provided in the following subsections. 

\begin{table*}[pos=tbp]
    \centering
    \caption{Programming Language Distribution of Source Repositories in ApacheCM}
    \label{tab:lang_stats}
    \begin{tabular}{l*{10}{c}}
        \toprule
        \textbf{Language} & Java & C++ & Scala & TypeScript & Python & Lua & Go & Rust & Erlang \\
        \midrule
        \textbf{Count}    & 33   & 6   & 3     & 2          & 2      & 1   & 1  & 1    & 1 \\
        \bottomrule
    \end{tabular}
\end{table*}

\subsection{Data Acquisition}
\label{subsec:dataqquistion}

ApacheCM is constructed based on open-source repositories from the Apache Software Foundation (ASF). It is mainly because that as a globally well-known nonprofit organization, ASF is dedicated to fostering high-quality open-source software through open, collaborative, and community-driven development. The standardized development workflows and consistently structured metadata of ASF projects provide strong guarantees for the quality and reliability of ApacheCM.

ApacheCM was built from the top 50 Apache repositories on GitHub, ranked by their star counts. These repositories include widely used projects (such as Superset, ECharts, and Spark) and span multiple mainstream programming languages, offering comprehensive data coverage for the CMG task. The distribution of programming languages is shown in Table 2. ApacheCM contains commit records from the main or master branches of these repositories since 2015, totaling 612,367 raw commit records. Each commit record is represented as the basic unit and is described by eight fields as below, including the code diff, commit message, repository name, commit SHA, author name, affected files, date, and lines of code.

\begin{itemize}
    \item \textbf{Diff}: The code changes introduced by the commit, stored in standard \texttt{.diff} file format, capturing modified files and line-level additions and deletions. This field serves as the primary input for CMG models, reflecting the specific code modifications made by developers in the commit.
    \item \textbf{Commit Message}: The original commit description authored by the developer, used as the ground truth in the dataset. To enhance text quality and consistency, preprocessing steps removed pull request number references (e.g., ``\#1234'') and retained only the first line of the commit message, discarding multi-paragraph descriptions or redundant sub-commit details introduced in merge commits.
    \item \textbf{Repo Full Name}: The complete name of the repository (e.g., \texttt{apache/spark}) associated with each commit.
    \item \textbf{Commit SHA}: The unique SHA generated by Git for each commit. This field facilitates reproducibility, source code localization, and alignment with other datasets.
    \item \textbf{Author Name}: The name of the developer who authored the commit. This field is primarily used to identify and filter commits generated by automated tools or bot accounts (e.g., those containing ``[bot]'' in the name)
    \item \textbf{Files}: A list of all file names affected by the commit, enabling subsequent analyses of the commit's impact scope, file distribution, and file-level features.
    \item \textbf{Date}: The timestamp when the commit was formally recorded in the version control system.
    \item \textbf{LoC}: The total number of code lines modified in the commit, calculated as the sum of added and deleted lines.
\end{itemize}

\subsection{Data Filtering}
\begin{figure}[pos=h]
    \centering
    \includegraphics[width=\linewidth]{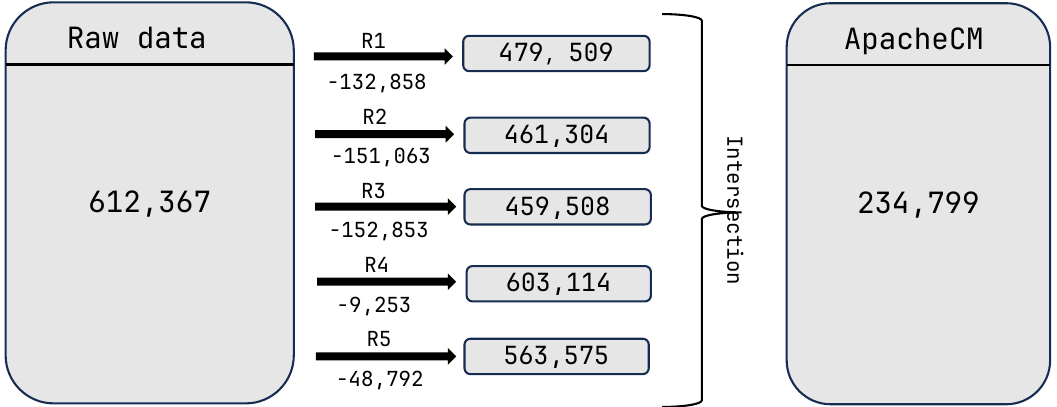}
    \caption{Commit Filtering Process and Results}
    \label{fig:filterstats}
\end{figure}

To ensure quality, usability, and adaptability for downstream tasks, the ApacheCM dataset was created with six filtering rules to exclude low-quality commit messages, as summarized in Table~\ref{tab:filtering_rules}. In total, 612,367 raw commit records were initially collected. Subsequently, multiple filtering rules were applied, with the number of commit records removed by each rule shown in Figure \ref{fig:filterstats}. After these steps, 234,799 commit records were included in ApacheCM.

\begin{table*}[pos=tbp]
    \centering
    \caption{Filtering Rules for ApacheCM}
    \label{tab:filtering_rules}
    \begin{tabular}{p{0.5cm} p{3cm} p{12.5cm}}
        \toprule
        \textbf{No.} & \textbf{Name} & \textbf{Description and Rationale} \\
        \midrule
        R1 & Message Length Filter & 
        \textbf{Description:} To filter out commit messages with fewer than 5 words or more than 50 words, based on space-separated word counts. \newline
        \textbf{Rationale:} Ensures a balance between conciseness and semantic completeness in natural language descriptions. \\
        \midrule
        R2 & Diff Length Filter & 
        \textbf{Description:} To remove commits with code diffs exceeding 300 lines. \newline
        \textbf{Rationale:} Eliminates large, coarse-grained commits that lack actionable specificity, focusing on manageable and interpretable changes. \\
        \midrule
        R3 & File Type Filter & 
        \textbf{Description:} To retain only commits that include modifications to at least one source code file in a mainstream programming language. \newline
        \textbf{Rationale:} Excludes commits involving only documentation or configuration, ensuring ApacheCM focuses on changes relevant to program semantics. \\
        \midrule
        R4 & Bot Filter & 
        \textbf{Description:} To exclude commits by authors whose names contain ``[bot]'', typically generated by automated tools or CI processes. \newline
        \textbf{Rationale:} Removes templated or repetitive content that lacks linguistic diversity and developer intent, improving ApacheCM's naturalness and representativeness. \\
        \midrule
        R5 & Revert \& Merge Filter & 
        \textbf{Description:} To filter out commits whose messages include keywords such as ``merge'' or ``revert''. \newline
        \textbf{Rationale:} Merge commits are version-control artifacts without functional semantics, and revert commits undo prior changes, both of which add noise and risk overfitting. \\
        \bottomrule
    \end{tabular}
\end{table*}

Finally, ApacheCM comprises 234,799 commit records collected from 50 high-quality Apache open-source projects, with each commit accompanied by comprehensive metadata, code diffs, and commit messages. The comparison of ApacheCM and existing datasets is shown in Table \ref{tab:datasets}.

In addition, ApacheCM includes statistical information on code diffs and commit message lengths (measured in token counts using a standardized tokenizer), as well as the average number of modified files and lines of code per commit. These statistics provide a clearer characterization of the dataset’s scope and distribution, as summarized in Table~\ref{tab:apachecm-stats}.

\begin{table}[pos=h]
  \caption{\textbf{Data Length and Change Size in ApacheCM}}
  \label{tab:apachecm-stats}
  \renewcommand{\arraystretch}{1.3}
  \begin{tabular}{l@{\hskip 3pt}r@{\hskip 4pt}r@{\hskip 4pt}r@{\hskip 2pt}r@{\hskip 2pt}r}
    \toprule
    \textbf{Type} & \textbf{Mean} & \textbf{Max} & \textbf{Median} & \textbf{\#Files} & \textbf{\#Lines} \\
    \midrule
      Code Diff    & 1089.88 & 54814.00 & 769.00  & 2 & 29 \\
      Message      & 15.77   & 109.00   & 15.00   & -- & -- \\
    \bottomrule
  \end{tabular}
\end{table}

\section{CoRaCMG Framework} 
\label{chap:methodology}
This section describes the methodology of our proposed three-phase framework, CoRaCMG. We first provide a brief overview of CoRaCMG and then give details of its three phases.

\subsection{Overview}
\begin{figure*}[pos=t]
    \centering
    \includegraphics[width=\linewidth]{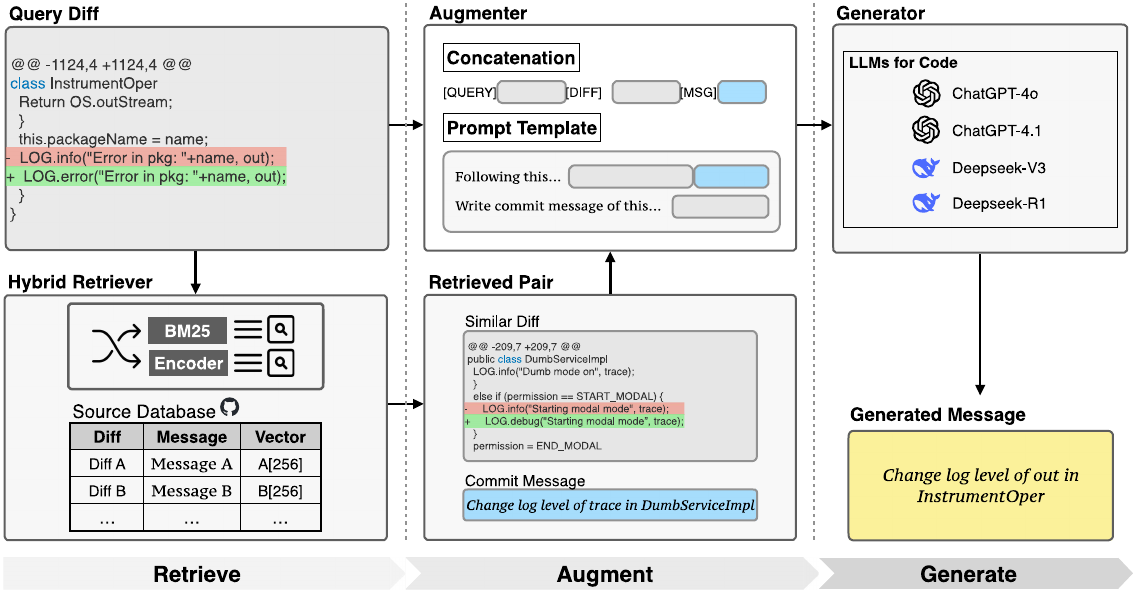}
    \caption{Overview of the CoRaCMG Framework}
    \label{fig:overview}
\end{figure*}

As shown in Figure \ref{fig:overview}, CoRaCMG comprises three phases, i.e., \textit{Retrieve}, \textit{Augment}, and \textit{Generate}, to generate the commit message for a specified code diff.  First, based on the query diff, CoRaCMG employs a \textsc{Hybrid Retriever} to select relevant example pairs from the ApacheCM-10K dataset. In this paper, we refer to the code diff that requires the generation of its corresponding commit message as the `\texttt{query diff}'. Additionally, we refer to the similar diff-message pair retrieved based on the query diff as the `\texttt{example pair}'. Then, both the query diff and the retrieved example pairs are sent to the \textsc{Augmenter}, in order to enrich the \texttt{query diff} with the contextual and semantic information from the example pairs. Finally, the augmented input, i.e., the query diff enhanced by the example pairs, is fed into the \textsc{Generator} to generate a commit message under the guidance of the example pairs.

\subsection{Phase I: Retrieve}
\label{sec:Retrieve}
Phase I intends to retrieve the relevant example pairs for the specified \texttt{query diff} to guide LLMs in generating commit messages. To more accurately measure the similarity between two code diffs and retrieve the most relevant example pair, we designed an advanced hybrid retriever that combines lexical and semantic similarity scores through weighted fusion. 

\textbf{At the lexical level}, we employ BM25~\citep{robertson2009probabilistic} to measure the relevance of a document to a given query diff by considering factors such as term frequency, document length, and term saturation. BM25 is a relevance scoring algorithm commonly used in information retrieval and search engines. It treats the query diff as a bag-of-words representation and computes lexical similarity scores between the query diff and each of the candidates.
 
\textbf{At the semantic level}, we employ a deep neural network-based code representation model to embed code diffs into fixed-dimensional dense vectors. Specifically, the Jina AI \texttt{jina-embeddings-v2-base-code} pretrained model $\phi(\cdot)$ is adopted to transform a code diff $d$ into a vector $v_d = \phi(d) \in \mathbb{R}^n$. This embedding captures the structural and semantic features of code fragments, surpassing superficial lexical matching to reflect the intent and logic of code changes. To optimize similarity computation efficiency, we apply unit vector normalization to dense vectors during generation. This step simplifies subsequent similarity calculations to efficient dot product operations, significantly enhancing retrieval efficiency. The resulting dense representations are stored alongside each commit record in ApacheCM-10K, serving as part of the reusable metadata available within ApacheCM-10K. After obtaining the embedded representations of code diffs, we computed their cosine similarity to measure semantic closeness.

Finally, following the normalization of the scores to a common scale, we combine the scores obtained from these two methods with equal weights (1:1) in our experiment, and then set it as the hybrid score. After calculating the hybrid scores between the query diff and all diffs in the ApacheCM-10K, the example pair with the highest score is retrieved as the most relevant pair. The hybrid retriever is expected to be more robust and effective compared to single retrievers that rely on only one similarity scoring method.

\textit{Data leakage issue:} It should be noted that due to the intersection between the ApacheCM-10K and the source database in the actual experiment, the example pair may be exactly the same as that of the ApacheCM-10K, resulting in data leakage issues. We append a simple mechanism to avoid this issue: when the retrieved diff is detected to be the same as the query diff, the pair with the second-highest hybrid score is selected to replace it.

\subsection{Phase II: Augment}
\begin{figure}[pos=h]
    \centering
    \includegraphics[width=\linewidth]{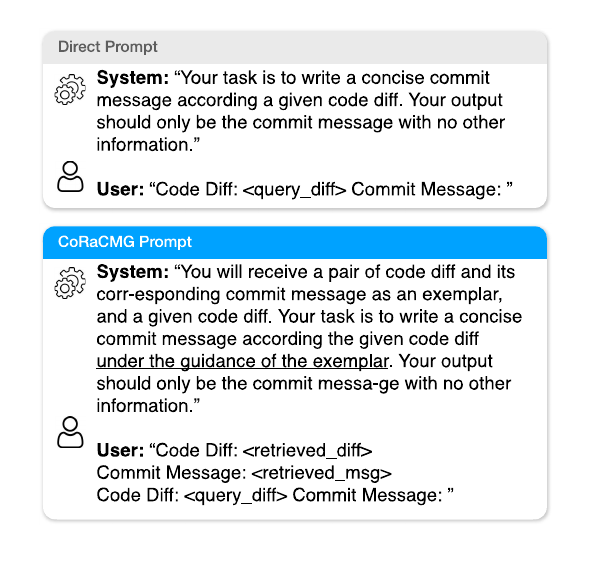}
    \caption{Direct prompt template and CoRaCMG prompt template}
    \label{fig:prompt}
\end{figure}

 Phase II aims at combining the query diff and the example pairs into specific formats to augment the input context. For ease of reference, we denote <\texttt{query diff}>, <\texttt{retrieved diff}>, <\texttt{retrieved msg}> as the query diff for generation, the retrieved code diff, and the corresponding retrieved commit message from the example pairs, respectively. To enable LLMs to comprehensively understand the semantics of the current commit while leveraging structural and stylistic information from example pairs, we propose a structured prompt template, as Figure ~\ref{fig:prompt} shows. This template explicitly incorporates three components: query diff, retrieved diff, and retrieved message. 

\subsection{Phase III: Generate}
In CoRaCMG, LLMs are employed to automatically generate commit messages. Specifically, the LLMs receive the augmented input from Phase II and generate the commit message guided by the example pair embedded within the input. Considering the robust contextual modeling and generalization capabilities of LLMs, they can directly generate commit messages without further fine-tuning. That is, by providing an example pair as input, LLMs will refer to this pair to generate commit messages. The more relevant the provided example pairs to the query diff, the more accurate the generated commit message will be.

\section{Experimental Setup}
\label{Experimental Setup}
In this section, we first define four research questions (RQs) to explore the performance of CoRaCMG. Then, the ApacheCM-10K is introduced, followed by the selected LLMs, and the metrics used to evaluate the experimental results are introduced. 

\subsection{Research Questions}
The \textbf{goal} of this research is to systematically evaluate the effectiveness of CoRaCMG in the CMG task by feeding augmented input with retrieved example pairs to LLMs to generate commit messages. For this purpose, we formulate the following four research questions (RQs): 

\textbf{RQ1: How do LLMs perform when they are directly applied to the CMG task?} This RQ aims to evaluate the performance of selected LLMs when they are applied directly to the CMG task. 
The answers to this RQ will serve as a baseline in the evaluation of CoRaCMG later.

\textbf{\{RQ2: What is the impact of varying retrieval configurations on the performance of CoRaCMG?} This RQ aims to investigate how different retrieval configurations affect the performance of CoRaCMG, with a particular focus on the variations in embedding models and weighting ratios in the \textsc{hybrid retriever}.

\textbf{RQ3:How does CoRaCMG perform in CMG by feeding one example pair to LLMs?} This RQ is designed to explore the performance of CoRaCMG by incorporating the most relevant example pair to generate the commit message for the query diff. The answer to this RQ would help evaluate CoRaCMG's performance across various objective and subjective metrics.

 \textbf{RQ4: Whether the number of example pairs used for input augmentation influences the quality of the commit messages generated by CoRaCMG?} This RQ explores how the number of example pairs, used as input exemplars, affects the quality of generated commit messages. The answer to this RQ would help optimize the input augmentation with a varying number of example pairs when using CoRaCMG in CMG.

Among these three RQs, RQ1 establishes the baseline to evaluate the performance of CoRaCMG, RQ2 intends to preliminarily evaluate CoRaCMG's performance in CMG, and RQ3 explores the influence of varying input augmentations on the performance of CoRaCMG.

\subsection{Experiment Dataset and Procedure}
\label{sec:dataset}
Considering the extensive size of the ApacheCM dataset constructed in Section~\ref{chap:ApacheCM}, we created a subset, \textbf{ApacheCM-10K}, which serves as the experimental dataset for CoRaCMG. For this purpose, multiple rounds of random sampling were conducted to select a total number of 10,000 commit records from the ApacheCM dataset, to ensure that ApacheCM-10K covers all nine programming languages. In this way, ApacheCM-10K preserves both the high quality of commit messages and the diversity of programming languages in the ApacheCM dataset. 

The experiments on ApacheCM-10K were executed to follow a complete generation process independently for each commit record. That is, example pairs are first selected from the retrieval candidate set and then used to generate the commit message for the current commit based on these pairs. Especially, during the process of constructing the retrieval candidate set, we strictly limited its scope to only include historical commits that come from the same project as the current query diff in ApacheCM-10K. This aims to maintain a uniform writing style and ensure context coherence between the query diff and example pairs, reflecting how developers actually work in real-world development scenarios, where they usually only look at historical commits within the same project.

Note that in CoRaCMG, the quality of example pairs directly influences the writing styles that LLMs can learn. Therefore, the use of well-filtered and preprocessed historical commits as the basis of retrieval is of critical importance for the evaluation of CoRaCMG.

\subsection{LLM Selection}
\label{sec:llm}
Table \ref{tab:model-classification} lists the LLMs selected for our experiments, including three open-source models and two closed-source models. Among them, \textit{DeepSeek R1} is the unique reasoning model. 
In this paper, the temperatures of all these LLMs are set to 0.0 to minimize randomness caused by LLMs in our experiment and ensure stable and deterministic outputs.

\begin{table*}[hb]
  \centering
  \caption{\textbf{LLMs Used for Generation}}
  \label{tab:model-classification}
  \renewcommand{\arraystretch}{1.3}
  \begin{tabular}{lccc}
    \toprule
    \textbf{Model Name} & \textbf{License Type} & \textbf{Reasoning Model} & \textbf{Release Date} \\
    \midrule
    GPT-4o       & Closed-source & No  & May 2024 \\
    GPT-4.1      & Closed-source & No  & April 2025 \\
    DeepSeek-V3  & Open-source   & No  & March 2025 \\
    DeepSeek-R1  & Open-source   & Yes & January 2025 \\
    Qwen-2.5     & Open-source   & No  & September 2024 \\
    \bottomrule
  \end{tabular}
\end{table*}

\subsection{Evaluation Design}
\label{sec:metrics}

The quality of LLM-generated commit messages will be evaluated from both objective and subjective perspectives. 

\subsubsection{Objective Evaluation}
\label{sec:objective_metrics}
Regarding objective evaluation, we not only introduce four typical metrics but also enhance the tokenization process.

\paragraph{Objective Metrics:} 
Following prior work on CMG~\citep{xu2019commit,hoang2020cc2vec,dong2022fira}, we employ the following four evaluation metrics to quantitatively assess the similarity between LLM-generated commit messages and human-written ones. Higher scores on these metrics indicate better performance.

\begin{itemize}
    \item \textbf{BLEU}~\citep{papineni2002bleu}: Originally used in machine translation, BLEU computes the n-gram precision between the generated message and the ground truth. We utilize Google BLEU to avoid the instability and excessive penalties for short sentences often seen with corpus-level BLEU, ensuring more reasonable evaluation at the sentence granularity.
    \item \textbf{Rouge-L}~\citep{lin2004rouge}: Standing for Recall-Oriented Understudy for Gisting Evaluation, this metric measures the longest common subsequence (LCS) between sentences, accounting for overlapped units such as n-grams and word sequences.
    \item \textbf{METEOR}~\citep{banerjee2005meteor}: This metric is calculated based on the harmonic mean of unigram precision and recall, with a higher weight on recall. It aligns well with human judgment and is widely used in translation tasks.
    \item \textbf{CIDEr}~\citep{lin2004rouge}: Originally developed for image captioning, CIDEr quantifies similarity based on the consensus between the generated message and the ground truth. It employs TF-IDF weighting to highlight salient n-grams, better capturing alignment with human-authored text.
\end{itemize}

\paragraph{Tokenization Enhancement:}
Standard tokenization strategies often struggle with the specific characteristics of commit messages, such as user-defined terminologies (e.g., camelCase), special symbols (e.g., ``\_'', ``/''), and mixed-case formats. Incorrect segmentation (e.g., treating ``bug-fix'' as a single token) can negatively impact the accuracy of the aforementioned metrics, particularly BLEU and CIDEr. To mitigate this and ensure a fair evaluation, we enhance the standard tokenizer (based on tokenizer\_13a) from the following three perspectives before calculating the scores of objective metrics:

\begin{itemize}
    \item \textbf{Symbol-Based Segmentation}: We introduce regular expression matching to identify non-alphanumeric characters (e.g., hyphens in ``bug-fix'') and insert spaces around them to enforce segmentation into independent tokens.
    \item \textbf{CamelCase Decomposition}: To handle prevalent camelCase naming (e.g., ``handleRequest''), we use regular expressions to identify word boundaries marked by uppercase letters, decomposing them into finer-grained semantic units.
    \item \textbf{Case Normalization}: After segmentation and decomposition, all tokens are converted to lowercase. This resolves mismatch failures arising from mixed-case usage (e.g., ``FIX'' vs. ``fix'') while preserving the semantic roles of the symbols handled in the previous steps.
\end{itemize}







\subsubsection{Subjective Evaluation}
\label{sec:subjective_metrics}

Considering subjective evaluation, we first explain the process of human evaluation and then introduce three typical subjective metrics.

\paragraph{Human Evaluation Process:}
To evaluate the semantic quality of generated commit messages, we recruited three individual participants with a background in Software Engineering and over three years of programming experience. \textcolor{black}{To ensure the objectivity and reliability of the user study, we carefully designed the evaluation procedure as follows:}

\begin{itemize}
    \item \textcolor{black}{\textbf{Blind Evaluation Setup}: First, participants were presented with the code diff. Then, they evaluated a shuffled mixture of commit messages. The mixture included the developer-authored reference message and the messages generated by different LLMs (both direct generation and CoRaCMG-enhanced).  The sources of the messages were strictly anonymized to prevent bias.}
    
    \item \textcolor{black}{\textbf{Evaluation Guidelines}: Before the formal evaluation, participants attended a tutorial session. We provided a unified scoring rubric with explicit examples for each score level (1 to 5) across the three metrics to ensure consistent evaluation standards.}
    
    \item \textcolor{black}{\textbf{Scoring and Aggregation}:} Each participant independently rated the same set of generated commit messages on a 1–5 scale according to the following three subjective criteria\textcolor{black}{. The final scores reported in our results are calculated by averaging the ratings from the three participants to minimize individual subjectivity.}
\end{itemize}

\paragraph{Subjective Metrics:}

Standard objective metrics often fail to capture the semantic nuances of generated messages and exhibit limited alignment with human judgment~\citep{Tao2022EmpiricalStudy}. \textcolor{black}{To complement the objective evaluation and comprehensively assess the generated messages, we adopted the following three subjective metrics:}

\begin{itemize}
    \item \textbf{Clarity}: Evaluates how easily the commit message can be understood, considering its wording, structure, and grammar.
    \item \textbf{Completeness}: Assesses how thoroughly the commit message captures all changes in the code diff, including important contextual details.
    \item \textbf{Correctness}: Assesses how accurately the commit message reflects the actual code changes, ensuring it avoids hallucinations or misinterpretations.
\end{itemize}

\section{Result and Analysis}
\label{chap: result}
\subsection{Using LLMs Directly for CMG (RQ1)}
This subsection aims to evaluate how LLMs perform when they are applied directly to the CMG task.
More specifically, the LLMs selected (see Section~\ref{sec:llm}) were applied to the ApacheCM-10K dataset (see Section~\ref{sec:dataset}) for the CMG task, without any input augmentation. In this situation, the input for the LLMs comprises exclusively of the essential task instruction and the query diff.

\subsubsection{Results of RQ1}
Table~\ref{tab:model-performance-directly} presents the performance of five LLMs selected in Section~\ref{sec:llm} and three SOTA baselines for the CMG task over four metrics introduced in Section~\ref{sec:metrics}. Compared with three SOTA baselines listed in the first three rows in Table~\ref{tab:model-performance-directly}, the five selected LLMs (in the fourth to eighth rows of Table~\ref{tab:model-performance-directly}) perform much better in CMG across all four objective metrics. Although CoRec outperformed the other two SOTA baselines across four metrics, these scores are still significantly lower than those achieved by any of the five LLMs.

When examining the performance of LLMs in detail, open-source models (DeepSeek models and Qwen-2.5) outperform GPT models in the CMG task. DeepSeek-V3 achieves the highest BLEU score of 10.03, while Qwen-2.5 achieves the highest CIDEr score of 8.36. Meanwhile, DeepSeek-R1 shows superior performance in semantic-sensitive metrics, with the highest Rouge-L score of 22.96 and the highest METEOR score of 20.56.

\begin{table}[pos=htbp]
\centering
\caption{Performance of Direct Use of LLMs vs. Baselines on ApacheCM-10K}
\label{tab:model-performance-directly}
\begin{tabular}{p{2cm}cccc}
\toprule
Model       & BLEU  & Rouge-L & METEOR & CIDEr \\
\midrule
NNGen \citep{liu2018neural}       & 2.83  & 5.38    & 5.48   & 0.70  \\
CoRec \citep{wang2021CoRec}       & 6.24  & 9.08    & 10.46  & 2.18  \\
CoreGen \citep{Nie2021CoreGen}     & 4.16  & 6.22    & 6.57   & 1.32  \\
\midrule
GPT-4o      & 9.12  & 21.08   & 18.51  & 7.12  \\
GPT-4.1     & 9.76  & 21.21   & 19.37  & 7.18  \\
DeepSeek-V3 & \textbf{10.03} & 22.75   & 19.64  & 8.18  \\
DeepSeek-R1 & 9.87  & \textbf{22.96}   & \textbf{20.56}  & 7.81  \\
Qwen-2.5    & 9.94  & 21.79   & 18.47  & \textbf{8.36}  \\
\bottomrule
\end{tabular}
\end{table}

\subsubsection{Analysis of Answer to RQ1}
In this subsection, the answer to RQ1 is analyzed and discussed from the following three perspectives.

\textbf{LLMs vs. SOTA Baselines:} As shown in Table~\ref{tab:model-performance-directly}, LLMs outperformed SOTA baselines in CMG across all four evaluation metrics. Regarding each metric, the highest score achieved by employing LLMs is at least twice as high as those attained by the SOTA baselines. This consistent superiority demonstrates the strong generalization and contextual comprehension abilities of LLMs, and enables LLMs to produce commit messages that are more accurate and more similar to commit messages written by human developers. Given these significant advantages over existing SOTA baselines, it is necessary to further explore how different LLMs perform in CMG.

\textbf{Open-source LLMs vs. Closed-source LLM:}
By zooming in on the LLMs from the perspective of the license types, it is observed that DeepSeek models and Qwen-2.5, standing as typical representatives of open-source LLMs, significantly outperform the typical closed-source model, i.e., GPT series, across four evaluation metrics. As shown in Table~\ref{tab:model-performance-directly}, DeepSeek-V3 achieves the highest score in BLEU (10.03), Qwen-2.5 achieves the highest score in CIDEr (8.36), while DeepSeek-R1 demonstrates the best performance in METEOR (20.56) and Rouge-L (22.96). Especially, the scores in the \textit{CIDEr} column of Table~\ref{tab:model-performance-directly} show the largest difference in LLMs' performance in CMG. Specifically, DeepSeek-V3 (8.18) and Qwen-2.5 (8.36) surpass GPT-4o (7.12) by 14.9\% and 17.4\%, respectively. This finding is highly encouraging, since the substantial score difference in CIDEr indicates that open-source models can generate higher-quality commit messages that incorporate user-defined terminologies and adopt writing styles similar to those of human developers. The results also reflect the rapid progress of the open-source community, where recent advances in LLM development have started to rival, or even outperform, the previously dominant closed-source models. 

\textbf{Reasoning Models vs. Standard Models:}
As shown in the last two rows of Table~\ref{tab:model-performance-directly}, the performance difference between the reasoning model DeepSeek-R1 and the standard model DeepSeek-V3 within the DeepSeek series is relatively small across all objective metrics. This result suggests that these two models are equally effective in the CMG task. This outcome aligns with our expectations. That is, the CMG task normally focuses on generating concise and accurate natural language descriptions based on the localized code context, rather than employing complex multi-step reasoning or long-chain inference. Therefore, enhanced reasoning abilities of LLMs may not lead to significant improvement in their performance in CMG. In addition, this observation implies that task-specific adaptability may potentially play a more important role than general reasoning ability in the context of CMG. Accordingly, the enhancement of reasoning ability may offer limited benefits to CMG.
 
\begin{tcolorbox}[colback=white,colframe=black,title=\textbf{Key Findings of RQ1:},fonttitle=\bfseries]
\begin{itemize}[leftmargin=0.5em]
    \item The performance of LLMs exhibits a significant superiority over that of the SOTA baselines in CMG. 
    \item The open-source LLMs (i.e., DeepSeek series models and Qwen-2.5) outperform the closed-source LLMs (i.e., GPT series models) when they are directly applied to CMG task.
    \item In the DeepSeek model series, the standard model (i.e., DeepSeek-V3) performs as well as the reasoning model (i.e., DeepSeek-R1) when they are directly applied to CMG task. 
   
\end{itemize}
\end{tcolorbox}

\subsection{Impact of Retrieval Configurations (RQ2)}\label{sec:RQ2}

\begin{table*}[pos=htbp]
  \centering
  \color{black}
  \caption{\textbf{\textcolor{black}{Performance of CoRaCMG under Varying Retrieval Configurations (RQ2)}}\\
  \footnotesize *Comparison of embedding models and weighting ratios between BM25 and Dense retrieval.}
  \label{tab:retrieval-ablation}
  \renewcommand{\arraystretch}{1.3}
  \begin{tabular}{lccccc}
    \toprule
    \textbf{Model} & \textbf{Weight (BM25:Dense)} & \textbf{BLEU} & \textbf{ROUGE-L} & \textbf{METEOR} & \textbf{CIDEr} \\
    \midrule
    CodeBERT & 10 : 0 \textcolor{black}{\textit{(without Dense)}} & 17.63 & \textbf{30.51} & 30.68 & 14.61 \\
             & 7 : 3  & 17.42 & 30.44 & 30.53 & 14.51 \\
             & 5 : 5  & 17.44 & 30.34 & 30.61 & 14.43 \\
             & 3 : 7  & \textbf{17.68} & 30.50 & 30.68 & \textbf{14.68} \\
             & 0 : 10 \textcolor{black}{\textit{(without BM25)}} & 16.92 & 29.81 & 29.42 & 13.99 \\
    \midrule
    Jina     & 10 : 0 \textcolor{black}{\textit{(without Dense)}} & 17.43 & 30.33 & 30.47 & 14.38 \\
             & 7 : 3  & 17.37 & 30.39 & 30.63 & 14.53 \\
             & 5 : 5  & 17.57 & 30.02 & \textbf{30.73} & 14.50 \\
             & 3 : 7  & 17.33 & 29.93 & 30.69 & 14.48 \\
             & 0 : 10 \textcolor{black}{\textit{(without BM25)}} & 17.43 & 29.96 & 30.63 & 14.33 \\
    \bottomrule
  \end{tabular}
\end{table*}

This subsection aims to investigate the impact of different retrieval configurations on selecting example pairs \textcolor{black}{and conduct an ablation study on the components of the hybrid retriever} in CoRaCMG. Specifically, we designed a set of comparative experiments focusing on two critical variables: the selection of embedding models (i.e., CodeBERT and Jina) and the weighting ratio that balances lexical and semantic retrieval scores. \textcolor{black}{To explicitly evaluate the performance impact of removing individual retrieval components, we conducted an ablation study by setting the weighting ratios to 10:0 (which removes the semantic dense component, labeled as \textit{without Dense}) and 0:10 (which removes the lexical sparse component, labeled as \textit{without BM25}).} Due to the limitations of time and computational resources, ApacheCM-1K, one subset of ApacheCM-10K, is constructed as the research dataset of RQ2 by randomly selecting 1,000 commit records from ApacheCM-10K. Furthermore, guided by the findings of RQ1, where both DeepSeek-V3 and DeepSeek-R1 exhibited superior performance with a marginal gap, DeepSeek-V3 is selected as the generator in this subsection's experiments to optimize computational efficiency. As a standard LLM, it provides an efficient baseline for conducting extensive experiments on various retrieval configurations. The experimental results will provide the foundation for identifying an effective and robust retrieval strategy adopted by the CoRaCMG framework.

\subsubsection{Results of RQ2}
Table \ref{tab:retrieval-ablation} presents the performance of CoRaCMG under various retrieval configurations, \textcolor{black}{serving a dual purpose: conducting an ablation study of the hybrid retriever and reporting scores for different combinations of embedding models and lexical-semantic retrieval weighting ratios.}

Regarding the weighting ratios, the metric scores fluctuate as the balance shifts between BM25 and dense vectors. For CodeBERT, the BLEU score is 17.63 at the pure lexical configuration (10:0) and reaches a peak of 17.68 at the 3:7 ratio, but decreases to 16.92 when using pure dense retrieval (0:10). A similar trend appears in the Jina model, where the METEOR score rises from 30.47 (10:0) to a peak of 30.73 (5:5), compared to 30.63 in the pure dense configuration (0:10).

In terms of embedding models, CodeBERT and Jina exhibit different peak performances across objective metrics. CodeBERT attains its highest BLEU score of 17.68 and CIDEr score of 14.68 at the 3:7 ratio. On the other hand, Jina achieves its highest METEOR score of 30.73 at the balanced 5:5 ratio. Furthermore, in the specific case of pure dense retrieval (0:10), Jina records a BLEU score of 17.43 and a Rouge-L score of 29.96, whereas CodeBERT records 16.92 and 29.81, respectively.

\subsubsection{Analysis of Answer to RQ2}
In this subsection, the answer to RQ2 is analyzed and discussed from the following three perspectives.

\textbf{Jina vs. CodeBERT:} Although CodeBERT achieves marginally higher peak scores on precision-oriented metrics such as BLEU and CIDEr under certain weighting configurations, Jina consistently demonstrates strong performance on semantic-sensitive metrics, particularly METEOR, and exhibits greater robustness across different retrieval configurations. Notably, under the pure dense retrieval configuration (0:10), Jina achieves a higher BLEU score (17.43) than CodeBERT (16.92), indicating that Jina embeddings are more self-sufficient in capturing semantic relevance without relying on lexical signals.
Beyond performance metrics, practical considerations further favor the adoption of Jina in the CoRaCMG framework. CodeBERT is constrained by a 512-token input limit, whereas Jina supports long-context inputs of up to 8,192 tokens. Given that code diffs in real-world repositories frequently exceed 512 tokens, the extended context capacity of Jina enables a more complete representation of code changes without truncation. Together, these observations suggest that Jina provides a more robust and practically applicable embedding choice for the CoRaCMG framework.

\textbf{Hybrid Retrieval vs. Single Mode Retrieval:} \textcolor{black}{The experimental results of single mode retrieval effectively serve as an ablation study for our framework. These ablation study results demonstrate that removing either the lexical or semantic component, thereby relying on a single mode, yields sub-optimal performance compared to the full hybrid configurations.} For example, when CodeBERT transitions from a hybrid configuration to pure dense retrieval \textcolor{black}{(\textit{without BM25})}, the BLEU score decreases noticeably from its peak value of 16.92. This trend suggests that while dense vectors effectively capture semantic intent, they may miss exact keyword matches, such as variable names or specific API calls, which are critical in commit messages. The incorporation of lexical retrieval (BM25) compensates for this limitation, ensuring that the retrieved example pairs are both semantically relevant and lexically accurate, thereby enabling the generation of higher-quality commit messages.

\textcolor{black}{\textbf{Lexical-Semantic Weighting Ratio:} With respect to weighting ratios, balanced configurations mostly achieve competitive or superior performance compared to heavily skewed configurations. Although CodeBERT with a 3:7 ratio achieves marginally higher scores in n-gram-based metrics like BLEU (17.68), the Jina model with a balanced 5:5 weighting ratio attains the highest METEOR score (30.73) among all evaluated configurations. This result is significant because METEOR incorporates recall and synonymy matching, offering a complementary perspective on semantic adequacy beyond purely n-gram-based precision. Furthermore, Jina exhibits superior practical applicability, as it supports long-context inputs of up to 8,192 tokens, whereas CodeBERT is constrained by a 512-token limit. This extended capacity ensures a more complete representation of large code diffs without truncation. Therefore, considering Jina's superior context length and its balanced performance at this ratio, the Jina embedding combined with a 5:5 weighting ratio is adopted as the standard configuration for subsequent experiments.}

\begin{tcolorbox}[colback=white,colframe=black,title=\textbf{Key Findings of RQ2:},fonttitle=\bfseries]
\begin{itemize}[leftmargin=0.5em]
    \item \textcolor{black}{The ablation study reveals that hybrid retrieval configurations mostly yield better and more stable performance than single mode retrieval}, mitigating the performance degradation observed in dense-only configurations.
    \item \textcolor{black}{Compared with CodeBERT, Jina demonstrates stronger semantic-oriented performance and greater robustness across different retrieval configurations.}
    \item \textcolor{black}{A balanced lexical-semantic weighting ratio provides an effective and robust trade-off.}
\end{itemize}
\end{tcolorbox}

\subsection{Performance of CoRaCMG with Single Retrieval Augmentation (RQ3)}
\label{sec:RQ3}
\textcolor{black}{This subsection investigates the performance of our proposed CoRaCMG when a single example pair retrieved from ApacheCM-10K is incorporated into the input. To further assess the semantic quality of the generated commit messages, we additionally conducted a human evaluation following the procedure described in Section~\ref{sec:subjective_metrics}. Specifically, we randomly sampled 370 instances from ApacheCM-10K for manual evaluation, with the sample size determined based on a 95\% confidence level and a 0.05 margin of error. These sampled instances were then evaluated by the three participants introduced earlier. For each instance, the participants assessed the generated commit messages according to the established human evaluation protocol.}


\begin{table*}[pos=tbp]
  \centering
  \caption{\textbf{Performance of Using LLMs Directly vs. Single Retrieval Augmented CoRaCMG}}
  \label{tab:performance-comparison}
  \renewcommand{\arraystretch}{1.3}
  \begin{tabular}{lccccc}
    \toprule
    \textbf{Model} & \textbf{Method} & \textbf{BLEU} & \textbf{Rouge-L} & \textbf{METEOR} & \textbf{CIDEr} \\
    \midrule
    GPT-4o      & Direct    & 9.12  & 21.08   & 18.51  & 7.12  \\
                & Enhanced  & 17.24 ($\uparrow$89\%) & 27.39 ($\uparrow$29\%) & 28.60 ($\uparrow$54\%) & 12.29 ($\uparrow$72\%) \\
    GPT-4.1     & Direct    & 9.76  & 21.21   & 19.37  & 7.18  \\
                & Enhanced  & 12.26 ($\uparrow$25\%) & 24.33 ($\uparrow$14\%) & 21.50 ($\uparrow$11\%) & 10.14 ($\uparrow$41\%) \\
    DeepSeek-V3 & Direct    & 10.03 & 22.75   & 19.64  & 8.18  \\
                & Enhanced  & 17.25 ($\uparrow$72\%) & 28.71 ($\uparrow$26\%) & 27.28 ($\uparrow$39\%) & 13.07 ($\uparrow$60\%) \\
    DeepSeek-R1 & Direct    & 9.87  & 22.96   & 20.56  & 7.81  \\
                & Enhanced  & \textbf{17.42} ($\uparrow$76\%) & \textbf{29.10} ($\uparrow$27\%) & \textbf{28.76} ($\uparrow$40\%) & \textbf{13.32} ($\uparrow$71\%) \\
    \textcolor{black}{Qwen-2.5}    & \textcolor{black}{Direct}    & \textcolor{black}{9.94}  & \textcolor{black}{21.79}   & \textcolor{black}{18.47}  & \textcolor{black}{8.36}  \\
                    & \textcolor{black}{Enhanced}  & \textcolor{black}{13.31 ($\uparrow$34\%)} & \textcolor{black}{25.53 ($\uparrow$17\%)} & \textcolor{black}{22.44 ($\uparrow$21\%)} & \textcolor{black}{11.53 ($\uparrow$38\%)} \\
    \midrule
    \textbf{Average Improvement} & - & 59.2\% & 22.6\% & 33\% & 56.4\% \\
    \bottomrule
    \multicolumn{5}{l}{Enhanced: Enhancing input context with a single retrieved example pair (CoRaCMG)}
  \end{tabular}
\end{table*}

\subsubsection{Results of RQ3}
\label{sec:answerRQ3}
\begin{figure*}[pos=htbp]
    \centering
    \includegraphics[width=0.8\linewidth]{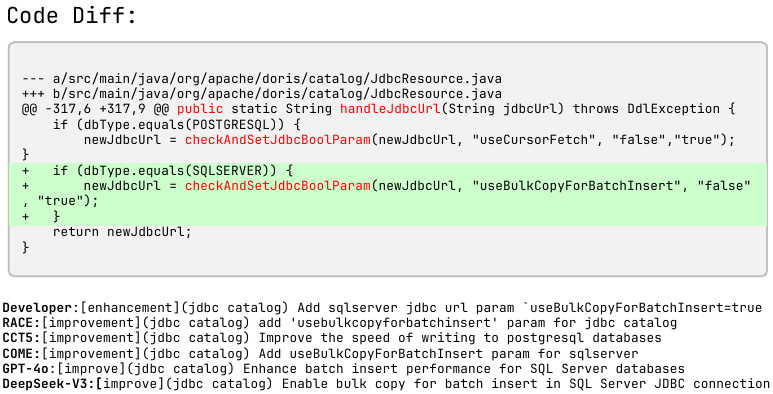}
    \caption{\textcolor{black}{An example of code diff and
its corresponding commit messages generated by the selected
models.}}
    \label{fig:messages_different_methods}
\end{figure*}

1) Objective Evaluation Results: Table~\ref{tab:performance-comparison} presents performance of our proposed CoRaCMG in which only the most relevant example pair of code diff and the corresponding message is fed into the LLMs to produce the commit message for the query diff. For each of the five LLMs, it is observed that compared to the direct use of the LLM, the performance of CoRaCMG with single retrieval augmentation in CMG is significantly enhanced across all four objective metrics. Figure~\ref{fig:messages_different_methods} shows an example of code diff and its corresponding commit messages generated by selected models. More specifically, using the enhanced DeepSeek-R1 in CoRaCMG achieves the highest scores of all four objective metrics, i.e., 17.42 in BLEU, 29.10 in Rouge-L, 28.76 in METEOR, and 13.32 in CIDEr. DeepSeek-V3 also demonstrated notable performance improvements, with the BLEU score experiencing a 72\% improvement and the CIDEr score showing a 60\% increase. Whereas, the enhanced GPT-4o in CoRaCMG brings the greatest increase in scores of four objective metrics. In contrast, GPT-4.1 and Qwen-2.5 yield relatively smaller jumps in these scores. The last row of Table~\ref{tab:performance-comparison} lists the average improvement ratio of each objective metrics when CoRaCMG employs single retrieval augmentation in CMG. It is found that the use of a single retrieved example pair in CoRaCMG leads to an average improvement of 59.2\% in BLEU, 22.6\% in Rouge-L, 33\% in METEOR, and 56.4\% in CIDEr.

\textcolor{black}{2) Subjective Evaluation Results: Table~\ref{tab:human} presents the results of the human evaluation conducted by three experienced participants. Consistent with the findings in the objective evaluation, our CoRaCMG framework (labeled as `Enhanced') outperforms the direct use of LLMs (`Direct') across all five models in terms of Clarity, Completeness, and Correctness. Among the evaluated models, DeepSeek-R1 equipped with CoRaCMG achieves the highest scores across all three subjective dimensions, recording 3.77 in Clarity, 3.22 in Completeness, and 3.81 in Correctness. In contrast, the `Reference' messages (i.e., ground truth written by developers) received notably lower scores, with 2.56 for Clarity, 1.55 for Completeness, and 2.78 for Correctness.} 
\begin{table}[!b]
    \centering
    \scriptsize  
   \color{black} 
    \caption{Results of Human Evaluation (Avg. of 3 participants)}
    \begin{tabular}{p{1.3cm}p{1cm}p{1.3cm}p{1.3cm}p{1.3cm}}
    \toprule
    \multicolumn{2}{c}{\textbf{Method}} & \textbf{Clarity} & \textbf{Completeness} & \textbf{Correctness}  \\
    \midrule
    \midrule
    \multicolumn{2}{c}{\textbf{Reference}} & 2.56 & 1.55 & 2.78 \\
    \midrule
    \multirow{2}{*}{\textbf{GPT-4o}} & Direct & 3.37 & 2.55 & 3.19 \\
    & Enhanced & 3.47 & 2.63 & 3.16 \\
    \midrule
    \multirow{2}{*}{\textbf{GPT-4.1}} & Direct & 3.55 & 2.81 & 3.57 \\
    & Enhanced & 3.62 & 2.89 & 3.61 \\
    \midrule
    \multirow{2}{*}{\textbf{DeepSeek-V3}} & Direct & 3.45 & 2.63 & 3.46 \\
    & Enhanced & 3.62 & 2.74 & 3.51 \\
    \midrule
    \multirow{2}{*}{\textbf{DeepSeek-R1}} & Direct & 3.54 & 2.90 & 3.65 \\
    & Enhanced & \textbf{3.77} & \textbf{3.22} & \textbf{3.81} \\
    \midrule
    \multirow{2}{*}{\textbf{Qwen-2.5}} & Direct & 3.23 & 2.34 & 3.02 \\
    & Enhanced & 3.23 & 2.35 & 3.03 \\
    \bottomrule
    \multicolumn{5}{p{8cm}}{Enhanced: Enhancing input context with a single retrieved example pair (CoRaCMG)}
    \end{tabular}
    \label{tab:human}
\end{table}

\subsubsection{Analysis of Answer to RQ3}
The results of RQ3 are analyzed and discussed from the following four aspects. 

\textbf{Performance of CoRaCMG in CMG varies in LLMs.}
In Table~\ref{tab:performance-comparison}, the rows labeled `Direct' and `Enhanced' compare the performance of each LLM before and after augmenting its input with the single retrieved example pair in CoRaCMG, respectively. Both the scores and the magnitude of score improvements vary when implementing different LLMs in CoRaCMG. This indicates that some LLMs (e.g., GPT-4o with the largest score jump) are sensitive to the code contextual retrieved augmentation, while others (e.g., GPT-4.1 and Qwen-2.5 with more moderate jumps) may not. Moreover, it is unexpected to observe that two prominent LLMs in the GPT family demonstrated significantly distinct improvements. 

\textbf{Using DeepSeek-R1 in CoRaCMG performed best in CMG.}
It is worth noting that, as the unique reasoning model in the selected LLMs, DeepSeek-R1 achieved the highest scores in all four objective metrics. The reason could be attributed to its unique mechanism for producing texts. This mechanism incorporates a type of internal reasoning that enables the model to more effectively identify and make use of the user-defined terminologies and writing styles presented in the example pair. These findings indicate that, when provided with enriched contextual inputs, reasoning models are better positioned to leverage the information, ultimately leading to the generation of more precise commit messages.

\textbf{Performance improvements based on CoRaCMG vary significantly across objective metrics.} 
As shown in the last row of Table~\ref{tab:performance-comparison}, the use of a single retrieved example pair in CoRaCMG leads to the highest average improvement of 59.2\% in BLEU and the lowest average improvement of 22.6\% in Rouge-L. Specifically, the average improvements of BLEU and CIDEr are approximately 60\%. The reason could be that BLEU prioritizes local n-gram precision, while CIDEr emphasizes TF-IDF–weighted term alignment. The retrieved example pair typically exhibits strong structural coherence and terminological accuracy, which aligns well with the evaluation principles of these two metrics. In contrast, the average improvements of Rouge-L and METEOR are relatively much lower. One possible reason is that Rouge-L and METEOR emphasize structural and semantic similarity, where the LLMs already perform well. As a result, the improvements by retrieving the similar example pair on these two metrics tend to be limited. 

\textbf{Performance improvements based on CoRaCMG are significant in human evaluation.} As indicated in Table~\ref{tab:human}, the commit messages generated by LLMs are mostly rated higher than the ground truths. First, all selected LLMs, regardless of whether they use retrieval augmentation, consistently outperform the ground truths. This disparity is most evident in the Completeness score (Reference: 1.55 vs. Enhanced DeepSeek-R1: 3.22). \textcolor{black}{To better understand their rating rationale, we collected feedback from the participants who reported that ground truths were frequently too brief or vague (e.g., using generic phrases like ``fix bug''), so they had to refer back to the code diffs to fully grasp the modifications. This observation practically corroborates that human-written messages often suffer from quality issues and a lack of essential information~\citep{maalej2010can, tian2022makes}. In contrast, LLMs generate informative and structured commit messages that comprehensively describe the code changes, thereby achieving higher ratings from human participants than the ground truth. Second, from the participants' perspective, our proposed CoRaCMG further enhances the quality of generated messages, particularly in terms of correctness and clarity, when compared to directly feeding code diffs into LLMs for message generation. They expressed that when code diffs are fed directly into LLMs for message generation, the models often produce generic descriptions in standard natural language, potentially deviating from the project's specific conventions. However, participants found that CoRaCMG effectively mitigated this issue. By incorporating a similar diff-message pair, CoRaCMG provides a reference for project-specific terminology and writing style. This guides the LLM to adopt the project-specific terminologies and writing styles consistent with the similar diff-message pair, thereby improving the quality of the generated messages and making them, as participants emphasized, more informative and practical.}


\textbf{How retrieved example pair impacts on the quality of generated commit messages?} We further analyzed the impact of the retrieved example pair on the quality improvement of commit messages generated by CoRaCMG from two perspectives: \textit{Terminology} and \textit{Writing Style}. Regarding \textit{Terminology}, commit messages often contain user-defined terms, such as API names, module identifiers, abbreviations, and other user-defined terms. When these terms are repeatedly used across historical commits, they constitute a consistent terminology system. By retrieving the most similar historical commit as the example pair, LLMs can reproduce such terms in the generated commit messages more accurately. This improves the consistency of terminologies and leads to improved performance of CoRaCMG on the objective metrics such as CIDEr and BLEU. As for \textit{Writing Style}, commit messages typically exhibit a highly templated and structured style, which implies a consistency that is especially pronounced within individual projects. By incorporating the retrieved example pair into the input context, LLMs implemented in CoRaCMG can learn and reuse this writing style to generate commit messages. This alignment enhances the conformity of the generated commit message with the project-specific writing conventions, thereby improving the METEOR and Rouge-L scores through an increased occurrence of shared subsequences with human-written messages.

\begin{table*}[tbp]
  \centering
  \color{black}
  \caption{\textbf{Performance Comparison of CoRaCMG against Multi-language Baselines on Java, C++, and Python}}
  \label{tab:multilang-performance}
  \renewcommand{\arraystretch}{1.25} 
  \resizebox{\textwidth}{!}{%
  \begin{tabular}{l|c|cccc|cccc|cccc}
    \toprule
    \multirow{2}{*}{\textbf{Model}} & \multirow{2}{*}{\textbf{Method}} & \multicolumn{4}{c|}{\textbf{Java}} & \multicolumn{4}{c|}{\textbf{C++}} & \multicolumn{4}{c}{\textbf{Python}} \\
    \cmidrule(lr){3-6} \cmidrule(lr){7-10} \cmidrule(lr){11-14}
     & & \textbf{BLEU} & \textbf{ROUGE-L} & \textbf{METEOR} & \textbf{CIDEr} & \textbf{BLEU} & \textbf{ROUGE-L} & \textbf{METEOR} & \textbf{CIDEr} & \textbf{BLEU} & \textbf{ROUGE-L} & \textbf{METEOR} & \textbf{CIDEr} \\
    \midrule
    RACE~\citep{shi2022race} & - & 12.25 & 21.06 & 20.60 & 8.27 & 9.18 & 13.80 & 18.36 & 2.91 & 8.23 & 16.15 & 14.76 & 4.00 \\
    COME~\citep{He2023COME} & - & 13.01 & \textbf{26.27} & 21.98 & \textbf{11.55} & 14.68 & 28.91 & 30.70 & \textbf{12.94} & 13.07 & 28.70 & 23.25 & 11.66 \\
    CCT5~\citep{Lin2023CCT5} & - & 12.93 & 23.45 & 23.41 & 11.31 & 14.36 & 27.57 & 30.05 & 11.17 & 14.80 & 26.66 & 22.56 & 10.49 \\
    \midrule
    \midrule
    \multirow{2}{*}{GPT-4o} 
      & Direct & 8.73 & 18.89 & 16.71 & 7.15 & 9.04 & 22.95 & 21.40 & 6.95 & 9.46 & 26.33 & 21.55 & 6.34 \\
      & Enhanced & \textbf{13.06} ($\uparrow$50\%) & 22.47 ($\uparrow$19\%) & 22.68 ($\uparrow$\textbf{36\%}) & 10.43 ($\uparrow$46\%) & 14.01 ($\uparrow$55\%) & 27.74 ($\uparrow$\textbf{21\%}) & 25.62 ($\uparrow$20\%) & 12.12 ($\uparrow$\textbf{74\%}) & 14.50 ($\uparrow$53\%) & 29.13 ($\uparrow$11\%) & 27.47 ($\uparrow$\textbf{27\%}) & 12.60 ($\uparrow$\textbf{99\%}) \\
    \midrule
    \multirow{2}{*}{GPT-4.1} 
      & Direct & 9.36 & 19.26 & 17.57 & 7.54 & 9.99 & 23.27 & 22.21 & 7.55 & 10.59 & 25.81 & 23.26 & 7.15 \\
      & Enhanced & 11.91 ($\uparrow$27\%) & 22.24 ($\uparrow$15\%) & 20.08 ($\uparrow$14\%) & 10.48 ($\uparrow$39\%) & 12.95 ($\uparrow$30\%) & 26.58 ($\uparrow$14\%) & 24.91 ($\uparrow$12\%) & 11.36 ($\uparrow$50\%) & 14.53 ($\uparrow$37\%) & 30.09 ($\uparrow$17\%) & 27.00 ($\uparrow$16\%) & 12.19 ($\uparrow$70\%) \\
    \midrule
    \multirow{2}{*}{DeepSeek-V3} 
      & Direct & 8.91 & 18.68 & 16.24 & 7.37 & 9.73 & 24.50 & 21.43 & 8.13 & 10.42 & 26.21 & 21.97 & 8.40 \\
      & Enhanced & 12.52 ($\uparrow$41\%) & 22.79 ($\uparrow$22\%) & 20.91 ($\uparrow$29\%) & 10.72 ($\uparrow$45\%) & 15.03 ($\uparrow$54\%) & 29.12 ($\uparrow$19\%) & 28.14 ($\uparrow$\textbf{31\%}) & 12.82 ($\uparrow$58\%) & 14.72 ($\uparrow$41\%) & 29.73 ($\uparrow$13\%) & 25.50 ($\uparrow$16\%) & 13.22 ($\uparrow$57\%) \\
    \midrule
    \multirow{2}{*}{DeepSeek-R1} 
      & Direct & 8.57 & 19.53 & 18.80 & 6.73 & 8.86 & 25.19 & 26.93 & 7.29 & 9.20 & 26.08 & 23.20 & 7.26 \\
      & Enhanced & 12.98 ($\uparrow$\textbf{51\%}) & 24.18 ($\uparrow$\textbf{24\%}) & \textbf{24.89} ($\uparrow$32\%) & 11.02 ($\uparrow$\textbf{64\%}) & \textbf{15.34} ($\uparrow$\textbf{73\%}) & \textbf{29.64} ($\uparrow$18\%) & \textbf{32.70} ($\uparrow$21\%) & 12.29 ($\uparrow$69\%) & \textbf{15.42} ($\uparrow$\textbf{68\%}) & \textbf{31.63} ($\uparrow$\textbf{21\%}) & \textbf{28.50} ($\uparrow$23\%) & \textbf{13.38} ($\uparrow$84\%) \\
    \midrule
    \multirow{2}{*}{Qwen-2.5} 
      & Direct & 9.26 & 18.70 & 16.07 & 7.79 & 9.55 & 22.41 & 19.89 & 7.94 & 9.70 & 25.76 & 20.90 & 7.16 \\
      & Enhanced & 11.73 ($\uparrow$27\%) & 21.49 ($\uparrow$15\%) & 18.82 ($\uparrow$17\%) & 10.12 ($\uparrow$30\%) & 12.06 ($\uparrow$26\%) & 26.53 ($\uparrow$18\%) & 22.27 ($\uparrow$12\%) & 10.48 ($\uparrow$32\%) & 12.53 ($\uparrow$29\%) & 28.08 ($\uparrow$9\%) & 23.57 ($\uparrow$13\%) & 10.70 ($\uparrow$49\%) \\
    \bottomrule
    \multicolumn{5}{l}{Enhanced: Enhancing input context with a single retrieved example pair (CoRaCMG)}
  \end{tabular}%
  }
\end{table*}

\textcolor{black}{\subsubsection{Generalizability across Programming Languages}}

\textcolor{black}{To evaluate the generalizability and robustness of CoRaCMG across different programming languages, we extended our evaluation to include Java, C++, and Python. Due to limitations of computational resources, we focused on these three languages as they are the most representative in the ApacheCM dataset. Specifically, we constructed language-specific test sets by partitioning the ApacheCM-10K dataset into subsets for these three languages. Additionally, we compared CoRaCMG against three additional SOTA baselines capable of handling multi-language CMG tasks: RACE~\citep{shi2022race}, COME~\citep{He2023COME}, and CCT5~\citep{Lin2023CCT5}.}

\textcolor{black}{Table~\ref{tab:multilang-performance} presents the performance comparison of CoRaCMG against these multi-language baselines and the direct use of LLMs across the three languages. The results indicate that CoRaCMG demonstrates consistent effectiveness and adaptability regardless of the programming language.}

\textcolor{black}{\textbf{Comparison with SOTA Baselines}: CoRaCMG enables LLMs to achieve performance that is comparable to, and in most cases, superior to SOTA baselines trained specifically for multi-language scenarios. In Python, the DeepSeek-R1 model enhanced by CoRaCMG achieves a BLEU score of 15.42 and a METEOR score of 28.50, significantly outperforming the best baseline CCT5, which achieved 14.80 and 22.56, respectively. In C++, the enhanced DeepSeek-R1 also establishes a new state-of-the-art with a BLEU score of 15.34, surpassing the strongest baseline COME (14.68).In Java, while the baselines remain competitive, GPT-4o enhanced by CoRaCMG achieves a BLEU score of 13.06, slightly edging out the best baseline COME (13.01).}

\textcolor{black}{\textbf{Impact of Retrieval Augmentation}: When comparing the ``Direct'' and ``Enhanced'' settings, retrieval augmentation consistently boosts performance across all languages and models. For instance, DeepSeek-R1 shows substantial relative improvements in BLEU scores across Java (51\%), C++ (73\%), and Python (68\%) when augmented with a single retrieved diff-message pair. Similarly, GPT-4o exhibits remarkable gains, particularly in Python, where the CIDEr score increases significantly by 99\% (from 6.34 to 12.60).}

\textcolor{black}{The consistent improvements across Java, C++, and Python can be attributed to the language-agnostic nature of CoRaCMG. CoRaCMG retrieves similar diff-message pairs regardless of the specific language syntax, thereby guiding LLMs to generate high-quality commit messages that adhere to the specific stylistic norms of each language. These findings confirm that CoRaCMG is not biased towards a specific language and possesses strong potential for multilingual applications.}

\begin{tcolorbox}[colback=white,colframe=black,title=\textbf{Key Findings of RQ3:},fonttitle=\bfseries]
\begin{itemize}[leftmargin=0.5em]
    \item  CoRaCMG demonstrates stable and substantial improvements in the CMG task by augmenting the input of LLMs with the retrieved example pair.
    \item \textcolor{black}{Human evaluation confirms that CoRaCMG outperforms both using LLMs directly to generate commit messages and the ground truths across subjective metrics.}
    \item When applying CoRaCMG in the CMG task, the reasoning model DeepSeek-R1 achieves the best performance.
    \item CoRaCMG achieves substantial quality improvements in generated commit messages by leveraging the example pair through reusing user-defined terminology and learning writing styles of historical commit messages. 
    \item \textcolor{black}{CoRaCMG exhibits robust generalizability across different programming languages (i.e., Java, C++, and Python), achieving performance that is comparable to, and in most cases superior to, specific multi-language baselines.}
\end{itemize}
\end{tcolorbox}

\subsection{Performance of CoRaCMG with Varying Retrieval Augmentation (RQ4)}
\label{sec:RQ4}
This subsection aims to further investigate the impact of the number of example pairs on the quality of commit messages generated by CoRaCMG. For this purpose, we designed a set of comparative experiments, varying in the number of example pairs (i.e., from one to five pairs), augmenting the input of LLMs used in CoRaCMG. \textcolor{black}{Due to the limitations of time and computational resources, we consistently employed the ApacheCM-1K dataset used in Section \ref{sec:RQ2}.} 
In addition, in this subsection, GPT-4o is selected as the LLM to be used in CoRaCMG, since it exhibited the greatest score improvements in four objective metrics, as reported in Section~\ref{sec:answerRQ3}.

\subsubsection{Results of of RQ4}
As illustrated in Figure~\ref{fig:metric_vs_examples}, the scores of all four objective metrics increase slightly when more example pairs are used to augment the input of GPT-4o in CoRaCMG. Specifically, when the number of example pairs grows from one to five, the METEOR score increases from 29.70 to 32.89, the Rouge-L score increases from 28.24 to 30.98, the BLEU score grows from 18.76 to 21.52, and the CIDEr score increases from 13.76 to 15.75. However, the increasing trend of scores seems to reach a plateau when more than three retrieved example pairs are fed to GPT-4o in CoRaCMG.

\begin{figure*}[pos=tbp]
    \centering
    \begin{tikzpicture}
    \begin{axis}[
        width=0.9\linewidth,
        height=8cm,
        xlabel={Number of Example Pairs},
        ylabel={Metric Score},
        ymin=10, ymax=35,
        xtick={1,2,3,4,5},
        ytick={10,15,20,25,30,35},
        legend style={at={(0.5,-0.2)}, anchor=north, legend columns=-1},
        legend cell align={left},
        grid=major,
        mark options={solid},
    ]
    
    \addplot+[blue, mark=*, thick] coordinates {
        (1,18.76) (2,20.0) (3,20.9) (4,21.1) (5,21.52)
    };
    \addlegendentry{BLEU}
    
    \addplot+[orange, mark=*, thick] coordinates {
        (1,28.24) (2,29.4) (3,30.5) (4,30.5) (5,30.98)
    };
    \addlegendentry{Rouge-L}
    
    \addplot+[green!60!black, mark=*, thick] coordinates {
        (1,29.7) (2,31.0) (3,32.1) (4,32.5) (5,32.89)
    };
    \addlegendentry{METEOR}
    
    \addplot+[red, mark=*, thick] coordinates {
        (1,13.76) (2,14.7) (3,15.28) (4,15.3) (5,15.75)
    };
    \addlegendentry{CIDEr}
    
    \end{axis}
    \end{tikzpicture}
    \caption{Performance of CoRaCMG by Using GPT-4o Across Different Numbers of Example Pairs}
    \label{fig:metric_vs_examples}
\end{figure*}


\subsubsection{Analysis of Answer to RQ4}
Figure~\ref{fig:metric_vs_examples} shows that the scores of four objective metrics are growing as the number of example pairs employed in CoRaCMG increases from 1 to 5. The reason could be that incorporating more example pairs to CoRaCMG enables the LLMs, taking GPT-4o as an example, to capture the terminological coherence and writing styles of commit messages in example pairs. Such an ability of LLMs, in turn, contributes to generating commit messages that are more informative and akin to those written by human developers. However, as the number of example pairs exceeds three, the magnitude of improvement begins to diminish. One possible reason could be the excessive input context, which introduces redundancy or even interference. This may distract the LLM and impair its focus on the most relevant code changes and their corresponding intent. 

In addition, the findings underscore the need to balance contextual richness and input length in practice when using CoRaCMG for the CMG task. On one hand, a moderate increase in example pairs can enhance the quality of generated commit messages. On the other hand, excessive example pairs may increase computational costs and result in performance saturation. Therefore, for our proposed CoRaCMG, three example pairs provide an optimal balance between effectiveness and efficiency in the CMG task. 

\begin{tcolorbox}[colback=white,colframe=black,title=\textbf{Key Findings of RQ4:},fonttitle=\bfseries]
\begin{itemize}[leftmargin=0.5em]
    \item Feeding more example pairs to GPT-4o consistently improves the quality of the commit messages generated by CoRaCMG.
    \item Retrieving more than three example pairs provides only marginal performance gains for CoRaCMG. 
\end{itemize}
\end{tcolorbox}

\textcolor{black}{\subsection{Qualitative Analysis of Failure Cases}}
\textcolor{black}{This subsection provides a qualitative analysis of cases where CoRaCMG produces low-scoring commit messages. For this purpose, we filtered the experimental results of DeepSeek-R1, the best-performing model in RQ3, to identify samples that scored zero on at least three objective metrics mentioned in Section~\ref{sec:objective_metrics}. From this filtered set, we randomly selected 220 distinct cases for manual inspection by following the sampling methodology mentioned in Section~\ref{sec:answerRQ3}. This sample size was determined based on a 95\% confidence level and a 0.05 margin of error.}

\textcolor{black}{Three participants carefully reviewed 220 failure cases, each of which includes the query diff, its reference commit message, the retrieved example pair for this query diff, and the commit message generated by CoRaCMG. 
After independently analyzing the potential causes of failure in each case, the participants held a group discussion to resolve discrepancies and reach a consensus.
Based on a systematic analysis, the causes contributing to the low-scoring cases are grouped into three main categories:}

\begin{figure}[pos=htbp]
    \centering
    \includegraphics[width=\linewidth]{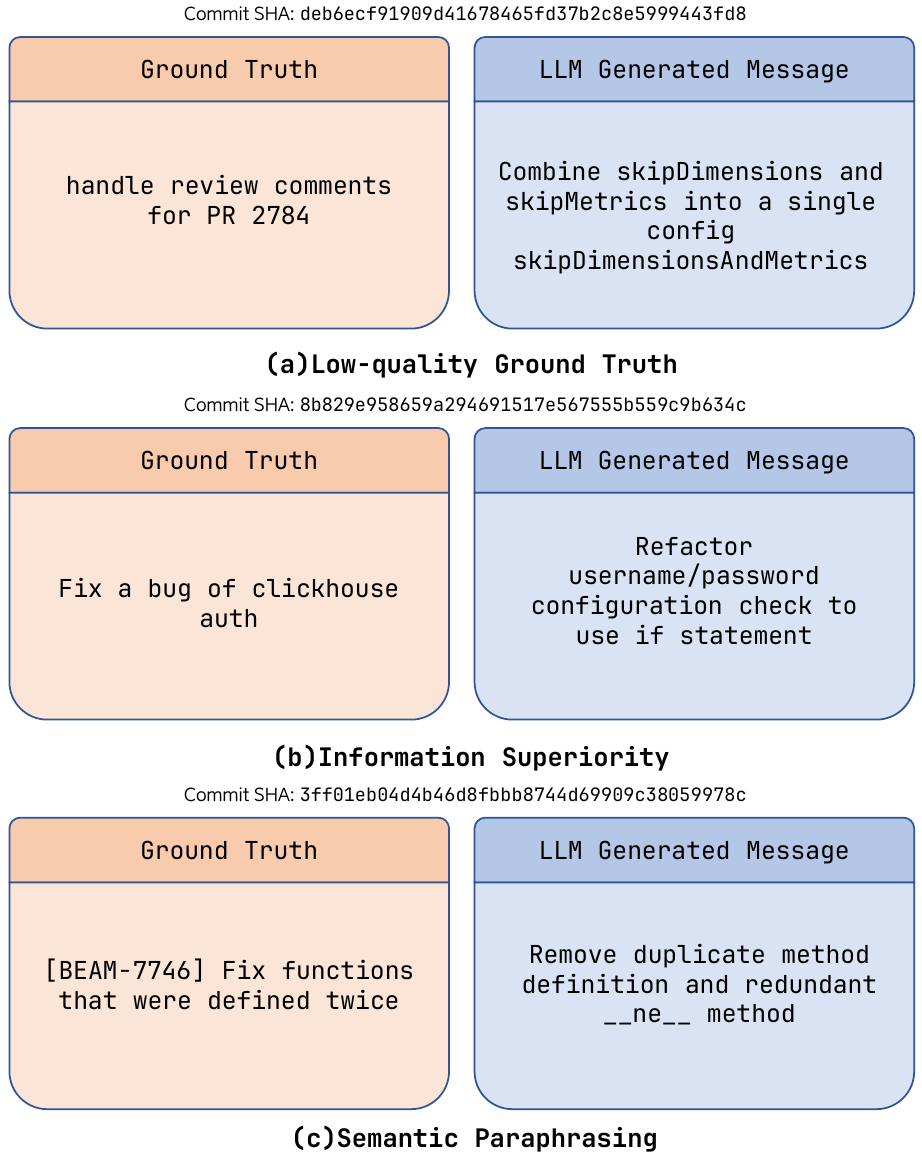}
    \caption{\textcolor{black}{Examples of Category I}}
    \label{fig:category_1}
\end{figure}

\subsubsection{Category I: Assessment Mismatch}
\textcolor{black}{In this category, the commit messages generated by the LLM are semantically reasonable and factually correct. However, they receive extremely low objective metrics scores due to the inherent limitations of the metrics or issues regarding the quality of the ground truths. (a)\textbf{Low-quality Ground Truth:} This scenario occurs when the ground truth itself is of poor quality or lacks substantive information (e.g., consisting solely of generic terms such as ``fix'', ``update'', and ``wip''). In contrast, CoRaCMG generates a descriptive message with the actual semantic content of the code diff. As illustrated in Figure~\ref{fig:category_1}(a), the ground truth is non-informative, whereas the generated message accurately describes the code diff. (b) \textbf{Information Superiority:} In these instances, the generated commit message provides more precise details than the ground truth. As shown in Figure~\ref{fig:category_1}(b), the LLM explicitly identifies the specific variables or logic modified, whereas the ground truth offers only a high-level summary. While the generated message is more informative, the lack of lexical overlap with the ground truth results in lower metric scores. (c) \textbf{Semantic Paraphrasing:} This occurs when the LLM conveys the same core intent as the ground truth but utilizes different vocabulary or syntactic structures. For example, in Figure~\ref{fig:category_1}(c), although the generated message is semantically equivalent to the ground truth, the divergence in lexical usage leads to low scores.}

\textcolor{black}{Distinct from the Assessment Mismatch described above, the following two categories represent scenarios where the generated messages genuinely deviate from the actual intent expressed in the code diffs.}

\subsubsection{Category II: Retrieval-Induced Interference}
\textcolor{black}{Unlike category I, where the generated messages were semantically valid, this category encompasses scenarios where the example pairs negatively influence the generation process, causing the LLM to produce factually incorrect or hallucinated messages. (1) Misleading Semantic Transfer: This phenomenon occurs when the retrieved diff exhibits high lexical overlap or structural similarity with the query diff, but serves a distinct purpose. The LLM fails to distinguish these semantic nuances and erroneously transfers the intent of the example pair to the generated message, resulting in an inappropriate message for the query diff. (2) Identifier Copying Error: In these instances, the LLM mimics the example pairs too mechanically. It directly copies specific identifiers—such as legacy Issue IDs, variable names, or class names—from the example pairs into the generated output, while ignoring the actual content of the query diff.}

\subsubsection{Category III: Source Context Limitations}
\textcolor{black}{This category attributes generation failures not to the LLM's reasoning capabilities, but to the inherent information deficit of code diff. In these cases, the code diff alone objectively lacks the sufficient context required to generate an accurate commit message. (1) Focus Misalignment: This scenario occurs when a single diff contains multiple, often unrelated, modifications (i.e., tangled changes). While the ground truth typically records only the primary modification, the LLM effectively becomes ``distracted'' by secondary code changes, causing the generated commit message to deviate from the core intent of code diff. (2) Contextual Deficiency: This refers to scenarios where the code diff is insufficient to infer the deeper rationale. For instance, code diffs related to high-level business rule updates or deprecation plans often require external knowledge—such as Issue discussions or Pull Request descriptions—to generate a valid and complete commit message. Without access to such external context, the LLM fails to accurately capture the underlying intent of code diff.}

\section{Threats to Validity}
In this section, we follow the guidelines provided by \cite{wohlin2012experimentation} to discuss the threats to the validity of this study from four perspectives, i.e., construct validity, internal validity, external validity, and conclusion validity.

\label{chap:threats}
\subsection{Internal Validity} 
Internal validity refers to potential biases or errors introduced by the experimental design. 
In this study, the objective evaluation of the performance of CoRaCMG heavily relies on four objective metrics, including BLEU, ROUGE-L, METEOR, and CIDEr. Although these metrics are widely used in text generation tasks, they are limited in capturing the semantics of commit messages and their alignment with the developers’ intentions. Especially, similarity-based metrics are commonly used in the CMG task; however, their reliability largely depends on the quality of the commit messages written by developers. In this work, ApacheCM is used as a comprehensive and high-quality dataset of commit messages to evaluate the performance of CoRaCMG in the CMG task. This partially mitigates the threat caused by the quality of commit messages in the research dataset. However, it is challenging to completely eliminate the subjectivity and unreliability inherent in metric-based evaluation.

Another potential threat comes from the LLM selected in the experiments to answer RQ4. In Section~\ref{sec:RQ4}, we selected GPT-4o to explore the impact of different numbers of example pairs on the performance of CoRaCMG. This is because GPT-4o exhibited the highest score improvement of all four metrics, as shown in Table~\ref{tab:performance-comparison}. However, as reported in Section~\ref{sec:answerRQ3}, DeepSeek-R1 achieved the highest scores of the objective metrics after incorporating example pairs. The results of RQ4 might be different if DeepSeek-R1 is used in CoRaCMG to answer RQ4. Meanwhile, a comparative study is expected to further investigate the impact of the number of example pairs on the performance of CoRaCMG in the CMG task. 


\subsection{External Validity} 
External validity concerns the generalizability of the findings reported in this study. 
Although ApacheCM was constructed to span a wide range of mainstream programming languages and high-quality open-source projects, it is predominantly based on repositories maintained by the ASF. This may limit the application of our findings to other development scenarios, such as proprietary enterprise systems or projects governed by different communities. However, the Apache community adheres to stringent standards and generates commit messages that are standardized in nature. Consequently, the performance of our proposed CoRaCMG may differ in cases where commit messages are collected from projects with less formal documentation. Another threat may be caused by the limited LLMs, i.e., five LLMs in three LLM series, that are used in this study to evaluate the performance of CoRaCMG. More popular LLMs are expected to be integrated into CoRaCMG for the CMG task. 

\textcolor{black}{Furthermore, CoRaCMG relies on the availability and relevance of historical data within the repository. In ``cold start'' scenarios where a repository is newly created or lacks sufficient history, the retrieval module may fail to identify relevant example pairs. In such cases, CoRaCMG effectively degrades to the standard zero-shot LLM approach (evaluated as the ``Direct'' setting in RQ1), which, while robust, lacks the project-specific knowledge. In addition, software projects often evolve over time, as coding guidelines or team composition change. If the retrieved example pairs are drawn from very early development phases, they may introduce obsolete writing styles of commit messages. While our hybrid retriever is designed to select the most semantically and lexically relevant example pairs of query diff, it currently does not explicitly penalize outdated retrieved example pairs. Future improvements of CoRaCMG could mitigate this by introducing temporal weighting to prioritize more recent commits during the phase of retrieving example pairs.}

\subsection{Construct Validity} 
Construct validity refers to the extent to which the experimental design accurately measures the theoretical constructs it is intended to evaluate. 
In this work, CoRaCMG incorporates contextual retrieval augmentation based on similar example pairs of code diff and the commit message, by combining sparse retrieval and dense retrieval with semantic encoders to ensure the quality of non-optimal selection of example pairs. This may reduce the quality of example pairs used to generate commit messages that are more precise and informative. Although we designed and implemented the tokenization enhancement (see Section~\ref{sec:metrics}) for CMG, the granularity of tokenization may still involve subjective decisions. Then, this could impact the consistency and reliability of objective metrics used in our work.


\subsection{Conclusion Validity}
The threats to conclusion validity concern the transparency and reproducibility of the experimental process. 
In this work, the experiments to evaluate the performance of CoRaCMG were detailed in Section~\ref{Experimental Setup}, including full release of source code and detailed configurations available at~\citep{replpack}. Meanwhile, we specified the implementation details such as the random seed used during dataset sampling, which collectively support the reproducibility of our study results. However, due to the limitations inherent in closed-source LLMs, our experimental procedure remains transparent and reproducible in all other aspects. 
    
\section{Conclusions and Future Work}
\label{chap: conclusions}
This paper proposed CoRaCMG, a Contextual Retrieval-augmented framework, to effectively improve the quality of generated commit messages. The evaluation of the performance of CoRaCMG was conducted by importing various LLMs and feeding retrieved example pairs to these LLMs. The experimental results reveal that CoRaCMG significantly improves the quality of commit messages generated by LLMs. These findings underscore the effectiveness of integrating contextual information, i.e., similar diff-message pairs that provide project-specific terminologies and writing styles, with LLMs. This integration offers a feasible and effective solution for the CMG task.

In the next steps, we plan to incorporate additional information beyond retrieving similar exemplars to further enhance the input of CoRaCMG. Taking issue-related commit messages as an example, the description and discussion details of the issues can be incorporated to help generate commit messages with higher quality.
First, it is expected to provide finer-grained retrieval strategies, such as local alignment based on function-level code fragments or change patterns, to improve semantic relevance and the transferability of retrieved example pairs. Second, more investigations on context enhancement are expected, such as combining structural information of source code (e.g., abstract syntax trees) with commit metadata, and incorporating additional sources such as pull request descriptions or issue discussions, etc. These richer contexts might help generate commit messages that are more accurate and valuable in practice.


\section*{Data availability}
We have shared the link to our replication package at~\citep{replpack}.

\section*{Acknowledgments}
This work was funded by the National Natural Science Foundation of China under Grant No. 92582203. The numerical calculations in this paper have been done on the supercomputing system in the Supercomputing Center of Wuhan University.

\printcredits


\bibliographystyle{cas-model2-names}

\bibliography{references}

\balance

\end{sloppypar}
\end{document}